\documentclass[11pt]{article} % use larger type; default would be 10pt

\usepackage{url}
\usepackage{hyperref}
\usepackage[utf8]{inputenc} % set input encoding (not needed with XeLaTeX)
\usepackage{authblk}

\usepackage{geometry} % to change the page dimensions
\usepackage{graphicx} % support the \includegraphics command and options
\usepackage{booktabs} % for much better looking tables
\usepackage{array} % for better arrays (eg matrices) in maths
\usepackage{paralist} % very flexible & customisable lists (eg. enumerate/itemize, etc.)
\usepackage{verbatim} % adds environment for commenting out blocks of text & for better verbatim
\usepackage{subfig} % make it possible to include more than one captioned figure/table in a single float
\usepackage{amsmath}
\usepackage{amssymb}
\usepackage{centernot}
\usepackage{mathtools}
\usepackage{caption}
\usepackage{fancyhdr} % This should be set AFTER setting up the page geometry
\usepackage{sectsty}
\allsectionsfont{\sffamily\mdseries\upshape} % (See the fntguide.pdf for font help)
\usepackage[nottoc,notlof,notlot]{tocbibind} % Put the bibliography in the ToC
\usepackage[titles,subfigure]{tocloft} % Alter the style of the Table of Contents

\title{Moments of characteristic polynomials and their derivatives for $SO(2N)$ and $USp(2N)$ and their application to one-level density in families of elliptic curve $L$-functions}
\author[1]{ I.A. Cooper}
\author[2]{ N.C. Snaith}
\affil[1]{School of Mathematics, University of Bristol, Bristol, BS8 3BU, UK}
\affil[2]{Corresponding Author n.c.snaith@bristol.ac.uk, School of Mathematics, University of Bristol, Bristol, BS8 3BU, UK}

\date{April 14, 2025} % Activate to display a given date or no date (if empty),
\begin{document}

\maketitle

\begin{abstract}
Using the ratios theorems, we calculate the leading order terms in $N$ for the following averages of the characteristic polynomial and its derivative:  $\left< \left|\Lambda_A(1 )\right| ^{r}  \frac{ \Lambda_A'(\mathrm{e}^{\mathrm{i} \phi}) }{ \Lambda_A(\mathrm{e}^{\mathrm{i} \phi})}  \right>_{SO(2N)}$ and $\left< \left|\Lambda_A(1 )\right| ^{r}  \frac{ \Lambda_A'(\mathrm{e}^{\mathrm{i} \phi}) }{ \Lambda_A(\mathrm{e}^{\mathrm{i} \phi})}  \right>_{USp(2N)}$. Our expression, derived for integer $r$,  permits analytic continuation in $r$ and we conjecture that this agrees with the above averages for non-integer exponents. We use this result to obtain an expression for the one level density of the `excised ensemble', a subensemble of $SO(2N)$, to next-to-leading order in $N$. We then present the analogous calculation for the one level density of quadratic twists of elliptic curve $L$-functions, taking into account a number theoretical bound on the central values of the $L$-functions. The method we use to calculate the above random matrix averages uses the contour integral form of the ratios theorems, which are a key tool in the growing literature on averages of characteristic polynomials and their derivatives, and as we evaluate the next-to-leading term for large matrix size $N$, this leads to some interesting multi-dimensional contour integrals, defined at the end of Section 2.1.3, which might be useful in other work.

\end{abstract}

\section{Introduction}

We define the characteristic polynomial 
$\Lambda_A(s)$ of $A \in U(N)$ as
\begin{equation}\label{1.1}
\Lambda_A(s) = \det ( I - s A^*) = \prod_{j=1}^N (1 - s e^{-i \theta_j}),
\end{equation}
with the eigenvalues of $A$ denoted by
$e^{i \theta_1}, \dots ,e^{i \theta_N},$ and $A^*$ being the conjugate transpose.  

Mixed moments for $U(N)$, in the spirit of  $\left< \left|\Lambda_A(1)\right| ^{k} \left| \Lambda_A'(1)  \right| ^{r} \right>_{A \in U(N)}$  (actually in some cases the results are for a variation of the characteristic polynomial that is multiplied by a factor so as to be real on the unit circle) have been studied extensively, starting with Hughes in 2001  \cite{kn:hug01}, by Dehaye \cite{kn:dehaye10,kn:dehaye10p}, by Winn \cite{kn:winn12} and Conrey, Rubinstein and Snaith in \cite{kn:crs06}, and more recently in two large collaborations \cite{kn:bbbcprs} and \cite{kn:basor_et_al18}.  These results have led to an understanding of the rate of growth of the leading order term when the matrix size $N$ is large, and several varied expressions for the leading order coefficient in terms of determinants, combinatorial sums, integrals or relating it to solutions of Painlev{\'e} equations. This has been extended to circular beta ensembles by Forrester \cite{kn:forrester22} and to higher derivatives in $U(N)$ by Keating and Wei \cite{kn:keawei1,kn:keawei2}, Barhoumi-Andr{\'e}ani \cite{kn:barhoumi} and Assiotis, Gunes, Keating and Wei \cite{kn:agkw} and generalised to $USp(2N)$, $SO(2N)$ and $O^-(2N)$ by Altu{\v g} et al \cite{kn:abprw14} and Andrade and Best \cite{kn:andbes}. There is an ongoing effort to give explicit expressions for the leading order coefficient also when the exponents $k$ and $r$ are non-integer.  Early results in this direction generalised to $k \in \mathbb{Z}, \ 2r \in \mathbb{Z}$  \cite{kn:winn12} or $k \in \mathbb{R}$, $r \in \mathbb{Z}$ \cite{kn:dehaye10p}, and in the past few years Assiotis, Keating and Warren \cite{kn:akw22} have extended both the exponent on the characteristic polynomial and its derivative to real numbers and expressed the leading order coefficient as the expectation value of a certain random variable, with an extremely recent extension of this to averages of products of many derivatives of any order raised to any positive real powers in \cite{kn:agkw}. Similar work has been carried out for other ensembles by Assiotis, Gunes and Soor in \cite{kn:ags22}.

Here, we calculate similar quantities over $SO(2N)$ and $USp(2N)$, with the distinction that for the purposes of our application to the one level density of zeros of $L$-functions, we do not always evaluate the characteristic polynomial at the point 1, as is the case in the aforementioned literature. In this way it is more akin to the $n$-correlation and $n$-level density calculations in \cite{kn:consna06,kn:consna07, kn:consna08, kn:massna16}. To start with, in Section \ref{sect:mixed_moment_even}, we shall focus on the expected value 
\begin{equation}\label{eq:themoment}
\left< \left|\Lambda_A(1 )\right| ^{r}  \frac{ \Lambda'(\mathrm{e}^{\mathrm{i} \phi}) }{ \Lambda(\mathrm{e}^{\mathrm{i} \phi})}  \right>_{SO(2N)},
\end{equation} (note that the equivalent quantity for $SO(2N+1)$ would always be zero due to the functional equation of the characteristic polynomial). The same method may also be applied to $USp(2N)$, and we have included the result of the equivalent calculation in Section  \ref{sect:mixed_moment_symp}.

An analogous  quantity, a joint average of the Riemann zeta function and its logarithmic derivative,  have been studied in the number theory literature for a slightly different purpose by Fazzari in \cite{kn:fazzari}.  Similar moments on the random matrix theory side have been considered by Simm and Wei \cite{kn:simmwei}.

In random matrix theory, the one level density gives us information about how the density of eigenvalues varies around the unit circle - about the likelihood of finding an eigenvalue in a particular interval on the unit circle if we draw a matrix at random from our ensemble.   If one could write $\left< \left|\Lambda_A(1 )\right| ^{r}  \frac{ \Lambda'(\mathrm{e}^{\mathrm{i} \phi}) }{ \Lambda(\mathrm{e}^{\mathrm{i} \phi})}  \right>_{SO(2N)}$ as an analytic expression in $r$, then we could use this quantity to calculate the one level density for the excised ensemble $\{A | A \in SO(2N), \Lambda_A(1) > \mathrm{e}^{\chi} \}$, which was introduced in \cite{kn:dhkms12} as a model for quadratic twist families of elliptic curve $L$-functions.  In Section \ref{sect:old_even} we make the assumption that our expression for (\ref{eq:themoment}) holds for complex $r$ with ${\rm Re}(r)>0$.

Although our motivation is number theoretical, we do not need to call on much number theoretical knowledge because the hard work has been done for us by Conrey, Farmer and Zirnbauer in \cite{kn:cfz2} who conjecture forms for averages of ratios of $L$-functions.  These expressions  have identical structure to their random matrix analogues.  So in Sections \ref{sect:Lratios} and \ref{sect:Lonelevel} we use these ratios conjectures to investigate the one-level density of $L$-functions associated with a family of elliptic curves, but we do so by simply mimicking the steps used in the random matrix calculation. However, for completeness we give a little introduction to the relevant number theory in the next section. 

%%%%%%%%%%%%%%%%%%%%%%%%%%%%%%%%%%%%%%%%%%%%
\subsection{Elliptic curve $L$-functions}

An elliptic curve $E$ may be described by an equation of the form $E: y^2 = x^3 + ax + b$, and a quadratic twist of $E$ by a fundamental discriminant $d$ would be the curve $E_d: d y^2 = x^3 + ax + b$. 

For such an elliptic curve, we can construct an $L$-function which has similar properties to the Riemann zeta function, for example an Euler product, a Dirichlet series, a Riemann Hypothesis. We write $L_E(s)$ for the $L$-function associated to $E$ and $L_E(s,\chi_d)$ for the $L$-function associated to $E_d$.  A theme in analytic number theory is that one can often deduce arithmetic information about number theoretical objects from properties of a related $L$-function. The Euler product of an $L$-function of an elliptic curve $E$ is
\begin{equation} \label{eq:eulerproductelliptic}
L_E(s) = \displaystyle \prod_{\text{bad \ primes}} \frac{1}{1 - \lambda(p) p^{ -s}}    \prod_{\text{good \ primes}}  \frac{1}{1 -\lambda(p) p^{ -s} + p^{ - 2s}},
\end{equation}
where $\lambda(p) := (p + 1 - N_p)/\sqrt{p}$, and $N_p$ counts solutions to the elliptic curve over the finite field of order $p$. This obeys the functional equation
\begin{equation}
L_E(s) = \omega(E) \left( \frac{2 \pi }{\sqrt{M}  } \right) ^{2s - 1} \frac{\Gamma(3/2 - s)}{\Gamma(1/2 + s) } L_E(1 -s).
\end{equation}
Here, the conductor $M$ is an integer associated to $E$, and the `good primes' are those which do not divide the conductor (\cite{kn:ashgross}, \cite{kn:rubsil02}). The Generalised Riemann Hypothesis suggests that zeros of such an $L$-function are constrained to lie on the critical line in the complex plane with real part 1/2.

%%%%%%%%%%%%%%%%%%%%%%%%%%%%%%%%%%%%%%%%%%%%%
\subsection{RMT and number theory}

Characteristic polynomials of unitary matrices have many features in common with $L$-functions.  According to the Katz-Sarnak philosophy, in some appropriate limit the zero statistics of a naturally-related family of $L$-functions tend to the equivalent statistics of eigenvalues from some subgroup of $U(N)$ \cite{kn:katzsarnak99a} \cite{kn:katzsarnak99b}. For unitary matrices, the analogue of the critical line is the unit circle, and the `Riemann hypothesis' is known to be true (as the eigenvalues are all constrained to lie on the unit circle). All of the zeros of the derivative lie inside the unit circle, which is equivalent to the hypothesis in number theory that all of the relevant zeros of derivatives of $L$-functions lie to the right of the critical line. For all subgroups of the unitary group, we have a functional equation, for example
\begin{equation}\label{eq:func_U}
\Lambda_A(s) = (-1)^N \text{det}(A) s^{N}\Lambda_{A^*}(s^{-1})
\end{equation}
for unitary matrices. Here, the appropriate Katz-Sarnak limit is the limit of large matrix size $N$, where $N$ plays the same role in the functional equation as the log of the conductor in the elliptic curve $L$-function case. The analogue of the critical value $L_E(1/2)$, where we are particularly interested in the value (or the order of vanishing) of the $L$-functions due to the Birch Swinnerton-Dyer conjecture, is the central value $\Lambda_A(1)$, which, as one would expect, is at the symmetry point of the functional equation. 

The excised model was proposed in \cite{kn:dhkms12} as a random matrix model for families of quadratic twists of elliptic curve $L$-functions which is applicable in the finite conductor regime. The one level density for the excised model was calculated in that paper using Gaudin's lemma. Here, we recalculate the one level density for the excised model using a method which requires the mixed moment
\begin{equation}
 \int _{SO(2N)}\frac{\Lambda(1)^r  {\Lambda' ( \mathrm{e}^{-\phi} )}}{\Lambda ( \mathrm{e}^{- \phi} )}  \mathrm{dA}.
\end{equation}
This is of interest, even though this calculation is already possible using an easier method, since the present method has an analogue in number theory - namely, via the ratios conjecture of \cite{kn:cfz2}, which involves averages (for example over a family) of ratios of $L$-functions. This allows us in principle to repeat the calculation in the number theory setting - e.g. calculating the one level density for zeros of elliptic curve $L$-functions in a family of quadratic twists - by simply following the same steps as in random matrix theory. 

Our method is based on that used in \cite{kn:hks}, where a conjecture for the one level density for quadratic twists of elliptic curve $L$-functions was derived. While this provided a very accurate conjecture for the one level density higher up the critical line, it failed to capture the behaviour of zeros close to the real axis. We expect that this is due to the approximations in the `recipe' of \cite{kn:cfz2} leading to a loss of information about the constraints on the central value of the $L$-functions in our family (\cite{kn:kohzag81},  \cite{kn:baruch_mao},\cite{kn:waldspurger81}), which force the central values either to take the value zero, or else obey
\begin{equation}
\label{eq:kz_bound}
L_E\left(\frac{1}{2}, \chi_{d}\right) > \kappa_{E}\left| d \right|^{-1/2}.
\end{equation}
For more discussion of how this bound should be interpreted in a random matrix model, see \cite{kn:dhkms12} and \cite{kn:coomorsna}. Our goal here is to use the calculation for the excised model as a template to conjecture an expression for the one level density for the $L$-functions. 

The means we have to do this, without calling too deeply on number theoretical knowledge, is the similarity of the following two formulae. 
Theorem 4.3 and Lemma 6.8 from \cite{kn:cfz2} tell us that
\begin{align}
& \int _{SO(2N)} \frac{\Lambda(1)^{K-1} \Lambda(\mathrm{e}^{-\alpha}) }{\Lambda ( \mathrm{e}^{-\gamma} )}  \mathrm{dA} 
 = \frac{ (-1)^{K(K-1)/2} 2^K}{(2 \pi \mathrm{i})^K K!} \mathrm{e}^{-N \alpha }\nonumber \\
&\qquad \times \oint\dots \oint   \exp \left( N\displaystyle \sum_{l=1} ^K w_l \right) \frac{ H^{SO(2N)}_{\alpha,\gamma}(w_1,\dots,w_K) \Delta ( w_1^2, \dots , w_K^2)^2 \prod_{i = 1} ^K w_i \mathrm{d}w_i  } { \displaystyle \prod_{n = 1} ^ K  (w_n - \alpha)(w_n + \alpha)   w_n ^ {2K-2}  },
\end{align}
and Conjecture 5.3 with Lemma 6.8 in \cite{kn:cfz2} gives us
\begin{align} 
   &\displaystyle \sum _{\substack{0 < d \leq X \\ \omega_E \chi_d( -M) = 1}} \left[  \frac{  L_E^{K-1} (1/2 , \chi_d) L_E (1/2 + \alpha, \chi_d)   }{L_E(1/2 +  \gamma, \chi_d)  }   \right] 
   %\nonumber \\& \qquad
   = \frac{(-1)^{K(K-1)/2} 2^K }{K! (2\pi \mathrm{i})^K)}  \displaystyle \sum_{\substack{ 0<d\leq X \\ \omega_E \chi_d( -M) = 1}}  \left(  \frac{M |d|^2 }{4 \pi^2 }\right)^{ \left(- \alpha/2 \right)}  \nonumber \\
&\times\oint\dots \oint   \left(  \frac{M |d|^2 }{4 \pi^2 }\right)^{ \tfrac{1}{2} \sum_{k = 1}^K \left( w_k  \right)}  \left(  \frac{H_{\alpha,\gamma}(w_1,\dots,w_K) \Delta(w_1^2,\dots,w_K^2)^2\prod_{i = 1} ^K w_i \mathrm{d}w_i }{\displaystyle \prod_{n = 1} ^ K  (w_n - \alpha)(w_n + \alpha)   w_n ^ {2K-2} } \right)  \nonumber \\
& \qquad+  \mathcal{O}\left(X^{1/2 + \epsilon}\right).
\end{align}
Here the contours enclose zero and $\pm\alpha$.  The components such as $z(x)$ and $H_{\alpha,\gamma}$ will be discussed when we come to use these formulae in later sections (see \eqref{ratios_mess} and \eqref{comb_sum} plus \eqref{nested_integrals}), but the point here is to notice that if $N$ is equated to $\tfrac{1}{2}\log(M|d|^2/(4\pi^2))$, then the structure of these two formula, one for an average over matrices from $SO(2N)$ with Haar measure, and one an average over a family of elliptic curve $L$-functions, is near identical. 

%%%%%%%%%%%%%%%%%%%%%%%%%%%%%%%%%%%%%%%%%%%%%%%
\section{A mixed moment for $SO(2N)$} \label{sect:mixed_moment_even}

We would like to calculate the following ratio of characteristic polynomials averaged over $SO(2N)$:
\begin{equation}\label{eq:momentintegral}
 \int _{SO(2N)}\frac{\Lambda(1)^r  {\Lambda' ( \mathrm{e}^{-\phi} )}}{\Lambda ( \mathrm{e}^{- \phi} )}  \mathrm{dA}.
\end{equation}
Note that as the eigenvalues of matrices $A\in SO(2N)$ occur in conjugate pairs, the characteristic polynomial has the form
\begin{equation}\label{eq:charpolySO2N}
\Lambda_A(s)=\prod_{j=1}^N (1-se^{-i\theta_j})(1-se^{i\theta_j}).
\end{equation}
We idealy want an expression for (\ref{eq:momentintegral}) valid for any complex value of $r$ with positive real part. In order to do so, we shall use the ratios theorem (which is only valid for integer $r$, but which eventually gives us an expression which can be analytically continued). 

The ratios theorem \cite{kn:cfz2}, in its contour integral statement, (see Theorem 4.3 and Lemma 6.8 from \cite{kn:cfz2}) tells us that
\begin{align}\label{ratios_mess}
\mathcal{R}_{SO(2N)}(\alpha,\gamma,K) :=& \int _{SO(2N)} \frac{\Lambda(1)^{K-1} \Lambda(\mathrm{e}^{-\alpha}) }{\Lambda ( \mathrm{e}^{-\gamma} )}  \mathrm{dA} \nonumber \\
 =& \frac{\mathrm{e}^{-N \alpha } (-1)^{K(K-1)/2} 2^K}{(2 \pi \mathrm{i})^K K!} \nonumber \\
\times \oint\dots \oint & \frac{  \displaystyle \prod_{1 \leq j < k \leq K} z (w_j + w_k) z (2 \gamma) \exp \left( N\displaystyle \sum_{l=1} ^K w_l \right) \Delta ( w_1^2, \dots , w_K^2)^2 \prod_{i = 1} ^K w_i \mathrm{d}w_i  } { \displaystyle \prod_{n = 1} ^ K z(w_n + \gamma)  (w_n - \alpha)(w_n + \alpha)   w_n ^ {2K-2}  },
\end{align}
where $z(x) = (1 - \mathrm{e}^{-x})^{-1} = \left(  \frac1x + \frac1{2} + \mathcal{O}\left(x\right)    \right)$, ${\rm Re}(\gamma)>0$, and the contours of integration enclose 0, $ \alpha$ and $-\alpha$. To avoid ambiguity, we shall take the contours to be nested, i.e. the $w_i$ contour shall completely enclose the $w_j$ contour whenever $j >i$.  The Vandermonde determinant is 
\begin{equation} \label{eq:vandermonde}
\Delta(x_1, \ldots, x_K)=\left|\begin{array}{ccccc}1&x_1&\cdots&x_1^{K-2}&x_1^{K-1}\\1&x_2&\cdots&x_2^{K-2}&x_2^{K-1}\\\vdots & \vdots & \ddots & \vdots & \vdots \\1&x_K&\cdots&x_K^{K-2}&x_K^{K-1}\end{array}\right|=\prod_{1\leq j<m\leq K} (x_m-x_j).
\end{equation}
%\centering
%\includegraphics[width=10cm]{contours1.png}
%\caption{An example of possible contours for two of the variables $w_i$ and $w_j$, where $j > i$.}
%\label{fig:cont}
%\end{figure}

Returning to (\ref{ratios_mess}), we shall define 
\begin{equation} \label{eq:C_even}
C^{SO(2N)}_{\alpha, \gamma, K} = \frac{ \mathrm{e}^{-N \alpha } (-1)^{K(K-1)/2} 2^K z(2 \gamma) }{(2 \pi \mathrm{i})^K  \Gamma(K+1)}.
\end{equation}
 First, we simplify the products slightly to obtain
\begin{align}\label{eq:presum}
\mathcal{R}_{SO(2N)}(\alpha,\gamma,K) = C^{SO(2N)}_{\alpha, \gamma, K} \oint \dots \oint \displaystyle \prod_{1 \leq j < k \leq K} z \left( w_j + w_k \right)   \Delta \left(w_1^2, \dots , w_K^2\right)^2 \nonumber \\
\times \displaystyle \prod_{n = 1} ^ K \left(\frac{  \exp \left( N  w_n \right) \left( w_n\right)^{3 - 2K} \mathrm{d} w_n } {  z\left(w_n + \gamma\right)  \left(w_n - \alpha\right)\left(w_n + \alpha \right)     }\right)  .
\end{align}
Note that we do not get poles from the $z(w_i + w_j)$ term due to cancellation with the Vandermonde $\Delta(w_1^2, \dots, w_K^2)$. Each contour has poles at $0$, $\alpha$ and $-\alpha$. We contract each contour onto the poles so that it consists of a small circle around each pole, with narrow necks connecting the circles.  The integral will cancel on each side of the neck as there are no branch points (note we are still assuming $K$ is an integer at this point). This leaves a sum in each variable over three circular contours:
\begin{align}\label{sum_even_ortho}
&\mathcal{R}_{SO(2N)}(\alpha,\gamma,K) = C^{SO(2N)}_{\alpha, \gamma, K} \displaystyle \sum _{\epsilon \in \{\pm \alpha,0  \}^K } \oint_{\epsilon_K} \dots \oint_{\epsilon_1} \displaystyle \prod_{1 \leq j < k \leq K} z \left(w_j + w_k\right) \Delta \left( w_1^2, \dots , w_K^2\right)^2  \nonumber \\
& \times \prod_{n = 1} ^ K\left( \frac{  \exp \left( N  w_n \right)   \left(w_n\right)^{3 - 2K} } {  z\left(w_n+ \gamma\right) \left(w_n- \alpha\right)\left(w_n+ \alpha \right)    }\mathrm{d} w_n\right),
\end{align}
where an integral with a subscript $\epsilon_j$ is the integral around a small circle centred on $\epsilon_j$, and $\epsilon = (\epsilon_1,\epsilon_2,\dots,\epsilon_K)$ is a $K$-tuple.

%%%%%%%%%%%%%%%%%%%%%%%%%%%%%%%%%%%%%%%%%%
\subsection{Evaluating the contributions to the sum }\label{sect:zero_contributions_even}

We are aiming to evaluate $\mathcal{R}_{SO(2N)}(\alpha,\gamma,K)$ for large values of $N$.  We will look at this asymptotically and also at the next-to-leading order term.  To perform the sum over $\epsilon$ in  (\ref{sum_even_ortho}) we need to consider separately terms where different numbers of variables are integrated round each of the three potential poles: $0$, $\alpha$ and $-\alpha$.  We show in the next section that in the case of a term where two or more of the $K$ variables are integrated around a pole away from 0, that term contributes zero to the sum for any value of $N$. If all variables are on contours around zero, that term will  be zero at leading order (see Section \ref{sect:all_zero_even}). This leaves us with evaluating the contribution when all variables except for one are evaluated around the pole at zero, which will be considered in Section \ref{sect:one_pole_even}.

%%%%%%%%%%%%%%%%%%%%%%%%%%%%%%%%%%%%%%%%%%
\subsubsection{Two or more variables integrated around $\pm \alpha$}\label{sect:geq_2_zeros}

Without loss of generality, we can choose $w_1$ and $w_2$ to be integration variables which we shall be integrating around either $\alpha$ or $-\alpha$. Consider the case where $\epsilon_1 = \epsilon_2 = \alpha$. The only singular parts of the integrand are the factors of $(w_1 - \alpha)$ and $(w_2 - \alpha)$ in the denominator.  Considering performing, say, the $w_1$ integral first, after evaluation of the residue at $w_1=\alpha$, this will leave us with factors of $(w_2-\alpha)$ from the Vandermonde in the numerator.  These exactly cancel out the $(w_2 - \alpha)$ in the denominator, so the integrand has no singularities and therefore no residues in $w_1$ or $w_2$. The same thing happens for $w_1 = w_2 = -\alpha$.

There are only two possible non-zero values for the $\epsilon_i$s, namely $\pm \alpha$, so if we have more than two non-zero $\epsilon_i$s then at least two of them must be equal and the integral will vanish.

Let us next consider the case where two of the $\epsilon_i$s are non-zero and non-equal. For the purposes of doing the integrals in $w_1$ and $w_2$, the other $w_i$s may be treated as constants. Let us consider only the parts of the integrand that either diverge or tend to zero as $w_1$ and $w_2$ approach $\pm  \alpha$. We shall show that this vanishes when we do the contour integrals in $w_1$ and $w_2$. This may be achieved by showing that
\begin{equation}\label{demo}
\oint_{-\alpha} \ \ \  \oint_{\alpha} \frac{ z(w_1 + w_2) (w_1 + w_2)^2 (w_1 - w_2)^2 } {(w_1 - \alpha)(w_1 + \alpha)(w_2 - \alpha)(w_2 +\alpha) } \mathrm{d}w_1 \mathrm{d}w_2
\end{equation}
vanishes. In order to do the integrals in (\ref{demo}), we evaluate them sequentially. Let us do the $w_1$ residue first, treating $w_2$ as a constant independent of $\alpha$:
\begin{align}
= &- 2 \pi \mathrm{i} \oint_{-\alpha}   \frac{ z(\alpha + w_2) (\alpha + w_2) (\alpha - w_2)} {(2 \alpha)}  \mathrm{d} w_2.
\end{align}
This has no poles in $w_2$, since the $(\alpha + w_2)^{-1}$ term from the series expansion of $z(\alpha + w_2)$ is cancelled out. This completes our demonstration that terms in (\ref{sum_even_ortho}) with two or more residues at evaluated at $\pm \alpha$ will have zero contribution to the total sum.

%%%%%%%%%%%%%%%%%%%%%%%%%%%%%%%%%%%%%%%%%%%%%%%%%%%%%%%%%%%%%%%%%%%
\subsubsection{All the poles at zero} \label{sect:all_zero_even}

We have established that terms with fewer than $K-1$ poles at zero do not contribute at all to the sum for any value of $N$. Let us now evaluate the contribution from the terms where all of the poles are integrated around zero, but this time we will scale the variables of integration by $1/N$ and look at the leading order term. We will find that even the leading order term of this calculation is negligible compared to the size of other terms in the sum (\ref{sum_even_ortho}) (see Section \ref{sect:one_pole_even}), but we can show that the coefficient of the leading order term here is zero, and as the method is useful later, we will demonstrate it in full detail.  We start with the terms in (\ref{sum_even_ortho}) where all poles are evaluated at zero:
\begin{align} \label{eq:allzeros1}
&C^{SO(2N)}_{\alpha, \gamma, K} \oint_{0} \dots \oint_{0} \displaystyle \prod_{1 \leq j < k \leq K} z \left(\frac{w_j}{N} + \frac{w_k}{N}\right) \exp \left( \displaystyle \sum_{l=1} ^K w_l \right) \Delta \left( \left(\frac{w_1}{N}\right)^2, \dots , \left(\frac{w_K}{N}\right)^2\right)^2  \nonumber \\
& \times \prod_{n = 1} ^ K\left( \frac{    \left(\frac{w_n}{N}\right)^{3 - 2K}\mathrm{d} \left(\frac{w_n}{N}\right) } {  z\left(\frac{w_n}{N}+ \gamma\right) \left(\frac{w_n}{N}- \alpha\right)\left(\frac{w_n}{N}+ \alpha \right)    }\right).
\end{align}
We can shrink the contours down to arbitrarily small circles around zero so that the $w_n$ variables are powers of $N$ smaller than $\alpha$ and $\gamma$. We can use $z(x) = (x)^{-1} + \mathcal{O}(1)$ to simplify the  $z$ factor in the numerator, and then have cancelation with the Vandermonde factor. 

We will just consider the leading order term, so we also now take the large $N$ limit to find that (\ref{eq:allzeros1}) equals
\begin{align}\label{eq:simplified_zero_poles}
&\frac{C^{SO(2N)}_{\alpha, \gamma, K}}{  z( \gamma)^{K} (-\alpha^2)^{K}}  \oint_{0} \dots \oint_{0} \displaystyle \prod_{1 \leq j < k \leq K}  \left(\frac{w_j}{N} - \frac{w_k}{N}\right) \left(\left(\frac{w_j}{N}\right)^2 - \left(\frac{w_k}{N}\right)^2\right)  \exp \left( \displaystyle \sum_{l=1} ^K w_l \right) \nonumber \\
& \times  \prod_{n = 1} ^ K \left(\frac{w_n}{N}\right)^{3 - 2K}  \mathrm{d} \left(\frac{w_n}{N}\right)\; \left(1+O(1/N)\right),
\end{align}
where the next-to-leading order term, which is one power of $N$ lower, comes only from the lower terms in the expansion of the $z(x)$ functions. 
%$ z \left(\frac{w_j}{N} + \frac{w_k}{N}\right)$ factors. 

Taking the $N$ dependence out of the integral, we get
\begin{eqnarray}
&&\frac{C^{SO(2N)}_{\alpha, \gamma, K} N^{\tfrac{1}{2}K^2-\tfrac{5}{2}K} }{  z( \gamma)^{K} (-\alpha^2)^{K}}   \\
&&\qquad\times \oint_{0} \dots \oint_{0}  \displaystyle \prod_{1 \leq j < k \leq K}  \left(w_j - w_k\right) \left(w_j^2 - w_k^2\right)  \exp \left( \displaystyle \sum_{l=1} ^K w_l \right)  \prod_{n = 1} ^ K w_n ^{3 - 2K}  \mathrm{d} w_n\; \left(1+O(1/N)\right). \nonumber
\end{eqnarray}
Note that 
\begin{equation}\label{eq:VDM}
\Delta(x_1,x_2, \dots , x_n) = \displaystyle \prod _{1 \leq j < k \leq n}  (x_j - x_k) = \displaystyle \sum _{\sigma \in S_{n}} \mathrm{sgn}(\sigma) x_1^{\sigma_0} x_2^{\sigma_1}\dots x_n^{\sigma_{n-1}},
\end{equation}
where $S_{n}$ is the permutation group of size $n$, where for convenience we have permutations of the numbers $0,1,2,\;\ldots,n-1$. Using (\ref{eq:VDM}),  we can rewrite the Vandermonde in terms of sums over the permutation group of size $K$:
\begin{eqnarray}\label{perms_even}
&&\frac{C^{SO(2N)}_{\alpha, \gamma, K} N^{\tfrac{1}{2}K^2-\tfrac{5}{2}K} }{  z( \gamma)^{K} (-\alpha^2)^{K}}  \oint_{0} \dots \oint_{0}\exp \left( \displaystyle \sum_{l = 1}^{K-1} w_l \right) \left( \displaystyle \sum_{\sigma \in S_{K}} \mathrm{sgn}(\sigma ) w_1^{2\sigma _0} w_2^{2\sigma _1} \dots w_{K}^{2\sigma _{K-1}}  \right) \\
&&\qquad\times    \left( \displaystyle \sum_{\rho \in S_{K}} \mathrm{sgn}(\rho ) w_1^{\rho_0} w_2^{\rho_1} \dots w_{K}^{\rho_{K-1}}  \right)  \prod_{n = 1}^{K}    w_n ^{3 - 2K} \mathrm{d} w_n \; \left(1+O(1/N)\right).\nonumber 
\end{eqnarray}
In each term of the sum over $\rho \in S_{K}$ in (\ref{perms_even}), let us relabel the variables of integration so that $w_1$ appears with exponent 0, $w_2$ with exponent 1 etc. This results in $K!$ identical terms, and the sign change from the relabelling cancels with the sign in the sum over $\sigma$. So, at leading order we have
\begin{align}\label{det_even}
&K! \frac{C^{SO(2N)}_{\alpha, \gamma, K} N^{\tfrac{1}{2}K^2-\tfrac{5}{2}K} }{  z( \gamma)^{K} (-\alpha^2)^{K}}   \oint_0 \dots \oint_0 \exp \left( \displaystyle \sum_{l = 1}^{K} w_l \right) \nonumber \\
&\times \left( \displaystyle \sum_{\sigma \in S_{K}} \mathrm{sgn}(\sigma )  w_1^{2\sigma_0} w_2^{2\sigma_1} \dots w_{K}^{2\sigma_{K-1}}  \right)  w_1^{3 - 2K} w_2^{4 - 2K} \dots w_{K}^{2 - K}  \mathrm{d}w_1 \dots \mathrm{d}w_{K} \nonumber \\
&= K! \frac{C^{SO(2N)}_{\alpha, \gamma, K} N^{\tfrac{1}{2}K^2-\tfrac{5}{2}K} }{  z( \gamma)^{K} (-\alpha^2)^{K}}  \oint_0 \dots \oint_0 \exp \left( \displaystyle \sum_{l = 1}^{K} w_l \right)\nonumber \\
&\times  \left|  \begin{array}{ccccc} 
1 & w_1^2 & w_1^4 & \dots & w_1^{2K-2} \\
1 & w_2^2 & w_2^4 & \dots & w_2^{2K-2} \\
\vdots & \vdots & \vdots & \ddots & \vdots \\
1 & w_{K}^2 & w_{K}^4 & \dots & w_{K}^{2K-2} \end{array} \right|  w_1^{3 - 2K} w_2^{4 - 2K} \dots w_{K}^{2-K}\mathrm{d}w_1 \dots \mathrm{d}w_{K}\nonumber  \\
&= K!\frac{C^{SO(2N)}_{\alpha, \gamma, K} N^{\tfrac{1}{2}K^2-\tfrac{5}{2}K}  }{  z( \gamma)^{K} (-\alpha^2)^{K}}   \nonumber \\
&\times   \left| \begin{array}{ccccc}
\oint \mathrm{e}^{w_1}w_1^{3 - 2K}\mathrm{d}w_1 & \oint \mathrm{e}^{w_1}w_1^{5 - 2K}\mathrm{d}w_1 & \dots & \oint \mathrm{e}^{w_1}w_1^{-1}\mathrm{d}w_1 & \oint \mathrm{e}^{w_1}w_1^{1}\mathrm{d}w_1 \\
\oint \mathrm{e}^{w_2}w_2^{4 - 2K}\mathrm{d}w_2 & \oint \mathrm{e}^{w_2}w_2^{6 - 2K}\mathrm{d}w_2 & \dots & \oint \mathrm{e}^{w_2}w_2^{0}\mathrm{d}w_2 & \oint \mathrm{e}^{w_1}w_1^{2}\mathrm{d}w_1 \\
\vdots & \vdots & \ddots & \vdots & \vdots \\
\oint \mathrm{e}^{w_{K}}w_{K}^{2 - K}\mathrm{d}w_{K} & \oint \mathrm{e}^{w_{K}}w_{K}^{4 - K}\mathrm{d}w_{K} & \dots & \oint \mathrm{e}^{w_{K}}w_{K}^{K - 2}\mathrm{d}w_{K} & \oint \mathrm{e}^{w_{K}}w_{K}^{K }\mathrm{d}w_{K} \end{array} \right| .
\end{align}
We can do the integrals inside the determinant using 
\begin{equation}\label{gamma_integral}
\frac{1}{\Gamma(z)} = \frac{1}{2 \pi \mathrm{i}} \int _C (-t)^{-z} \mathrm{e}^{-t} (-\mathrm{d}t)
\end{equation}
where the contour of integration comes in from $+\infty$, encircles the origin in the anticlockwise direction and then returns to $+\infty$. The exponent on $t$ will always be an integer in our case, so the integral on the two strands to $+\infty$ will cancel, leaving just an anticlockwise circle around the origin. So, (\ref{det_even}) becomes
\begin{align} \label{eq:zero_det_even}
K!\frac{C^{SO(2N)}_{\alpha, \gamma, K} N^{\tfrac{1}{2}K^2-\tfrac{5}{2}K} }{  z( \gamma)^{K} (-\alpha^2)^{K}}  \left| \begin{array}{cccccc} \frac{1}{\Gamma(2K-3)} & \frac{1}{\Gamma(2K-5)} & \dots & \frac{1}{\Gamma(3)} & \frac{1}{\Gamma(1)} & \frac{1}{\Gamma(-1)}\\
\frac{1}{\Gamma(2K-4)} & \frac{1}{\Gamma(2K-6)} & \dots & \frac{1}{\Gamma(2)} & \frac{1}{\Gamma(0)} & \frac{1}{\Gamma(-2)} \\
\vdots & \vdots & \ddots &  \vdots &\vdots & \vdots   \\
\frac{1}{\Gamma(K-2)} & \frac{1}{\Gamma(K-4)} & \dots & \frac{1}{\Gamma(4-K)} & \frac{1}{\Gamma(2-K)}  & \frac{1}{\Gamma(- K)}  \end{array} \right| ,
\end{align}
and this is the leading order contribution from the term in (\ref{sum_even_ortho}) where all residues are evaluated at zero.  Since $1/\Gamma(n) = 0$ when $n$ is a negative integer, the final column consists entirely of zeros, which means the determinant is zero.

%REMOVE??????????? If we repeat this calculation for the next-to-leading-order term in $N$, we get a sum of terms proportional to a similar %determinant to (\ref{eq:zero_det_even}), but with the arguments of the gamma functions in one of the rows decreased by one (this method shall be %applied explicitly in Section \ref{sect:one_pole_even}). All of these determinants are also zero, since all of the final columns consist entirely of zeros. 

%%%%%%%%%%%%%%%%%%%%%%%%%%%%%%%%%%%%%%%%%%%%%%%%%%%%%%%%%%%%%%%%
\subsubsection{Setting up the case of one pole away from zero}\label{sect:one_pole_even}

We shall consider the case where exactly one of the integrals is integrated around $\pm  \alpha$, and the rest are integrated around zero. Let us first integrate $w_1,\dots, w_{K-1}$ around zero and $w_K$ around $\alpha$, and call this term $\mathcal{T}^{SO(2N)}_1$.  There will be $K$ other terms exactly equivalent to this term (i.e. identical up to relabelling of the integration variables) contributing to our final result, and $K$ terms equivalent to to integrating  $w_1,\dots, w_{K-1}$ around zero and $w_K$ around $-\alpha$. If we call the latter $\mathcal{T}^{SO(2N)}_2$, then the total contribution to $\mathcal{R}_{SO(2N)}(K,\alpha,\gamma)$ from the terms with one pole away from zero will be $K \mathcal{T}^{SO(2N)}_1 + K \mathcal{T}^{SO(2N)}_2$. Let us find $\mathcal{T}^{SO(2N)}_1$ first, where we shall start by separating the $w_K$ variable from the products:
\begin{align}\label{scaled_sum_even}
\mathcal{T}^{SO(2N)}_1 &=   C^{SO(2N)}_{\alpha, \gamma, K}   \oint_{ \alpha} \oint_0\dots \oint_0  \displaystyle \prod_{1 \leq j < k \leq K} z \left( w_j + w_k\right) \;\Delta \left( w_1^2, \dots , w_K^2\right)^2 \nonumber \\
&\times \prod_{n = 1} ^ K \left(\frac{   \exp (Nw_n) w_n^{3 - 2K} } {  z\left(w_n+ \gamma\right) \left(w_n - \alpha\right)\left(w_n + \alpha \right)    } \right) \;\; \mathrm{d}  w_1\cdots\mathrm{d}  w_{K}\\
&= C^{SO(2N)}_{\alpha, \gamma, K} \oint_0\dots \oint_0 \displaystyle \prod_{1 \leq j < k \leq K-1} z \left( w_j+ w_k \right)   \;  \Delta \left( w_1^2, \dots ,w_{K-1}^2\right)^2     \nonumber \\
&\times \prod_{n=1}^{K-1}\frac{\exp(Nw_n) w_n^{3-2K}}{z(w_n+\gamma) (w_n-\alpha)(w_n+\alpha)}\nonumber \\
&\times\left( \oint_\alpha \left[ \prod_{j = 1} ^ {K-1}z(w_j+ w_K ) (w_K^2 -w_j ^2)^2 \right] \frac{\exp(Nw_K) w_K^{3 - 2K} }{z(w_K+\gamma) (w_K-\alpha)(w_K+\alpha)}\mathrm{d}  w_K\right) \mathrm{d}  w_1\cdots\mathrm{d}  w_{K-1} .\nonumber 
\end{align}
To continue, we integrate out $w_K$ (noting that the contour encircles only the pole at $\alpha$, making this a simple residue calcuation) and scale the remaining variables by $1/N$ (observing that the powers of $N$ from the Vandermonde cancel out with the factors from \newline $\prod_{n=1}^{K-1}(w_n/N)^{3-2K} \mathrm{d}(w_n/N)$):
\begin{align}\label{eq:rem}
\mathcal{T}^{SO(2N)}_1 &= C_{\alpha,\gamma,K}^{SO(2N)}  (2 \pi \mathrm{i})  \frac{ \mathrm{e}^{N\alpha} \alpha^{3 - 2K} }{z\left( \alpha + \gamma\right) (2\alpha)} \nonumber \\
\times &\oint _0 \dots \oint_0        \displaystyle \prod_{1 \leq j < k \leq K-1} z\left(  \frac{w_j}N + \frac{w_k}N \right)\Delta \left( w_1^2, \dots , w_{K-1}^2\right)^2  \nonumber \\
\times& \displaystyle \prod_{n=1}^{K-1}  \left[   z\left( \frac{w_n}{N} +\alpha \right)\left(\alpha^2 - \left(\frac{w_n}{N}\right)^2\right)^2     \frac{ \exp(w_n) w_n ^{3 - 2K} \mathrm{d}w_n }{z\left( \frac{w_n}N + \gamma \right) \left(  \frac{w_n}{N} - \alpha \right)\left(  \frac{w_n}{N} + \alpha \right)    }  \right].
\end{align}
We see that the $(\tfrac{w_n}{N}\pm \alpha)$ factors from the denominator cancel one of the similar factors in the numerator. 

From here, we shall keep track of both the leading order and the next-to-leading order term in $N$. Let us expand all the factors which have a $1/N$ term. To do so, we shall use
\begin{align}\label{eq:z1}
z\left( \frac x N\right) = N\left(  \frac1x + \frac1{2N} + \frac{x}{12N^2}    \right) + \mathcal{O}\left(N^{-2}\right) \\
\label{eq:z2} z\left( \alpha  + \frac x N \right)  = z\left( \alpha \right) \left( 1 + z(-\alpha) \frac x N    \right)      + \mathcal{O}\left(N^{-2}\right)   \\
\label{eq:z3} \frac 1 {z\left( \gamma + \frac x N    \right)} = \frac 1 {z(\gamma)}  \left( 1 - z(-\gamma) \frac x N   \right)  + \mathcal{O}\left(N^{-2} \right) \\
\label{eq:z4} \left(\frac{w_n}{N}\right)^2 - \alpha^2= -\alpha^2 + \mathcal{O}(N^{-2}).
\end{align}
Using these, (\ref{scaled_sum_even}) becomes
\begin{align}
\mathcal{T}^{SO(2N)}_1 &=  N^{(K-1)(K-2)/2}   C_{\alpha,\gamma,K}^{SO(2N)}  (\pi \mathrm{i})   \frac{\mathrm{e}^{N\alpha} \alpha^{2 - 2K}}{z\left( \alpha + \gamma\right)  }   \nonumber \\
&\times  \oint _0 \dots \oint_0     \displaystyle \prod_{1 \leq j < k \leq K-1}\left[ \frac 1 {w_j + w_k}  + \frac 1 {2N} +\mathcal{O}\left(N^{-2} \right)    \right]\Delta \left( w_1^2, \dots , w_{K-1}^2\right)^2  \nonumber \\
&\times\exp\left(\displaystyle \sum_{l = 1} ^{K-1} w_l \right)  \displaystyle \prod_{n=1}^{K-1}     \left[ \left( z(\alpha) + \frac{w_n} N z(\alpha)z(-\alpha) \right) \left(-\alpha^{2}  \right)    \frac 1 {z(\gamma)} \right.\nonumber \\
 &\times\left.  \left( 1 - \frac{w_n} N z(-\gamma)   \right)    w_n^{3 - 2K} \mathrm{d} w_n \left(1 + \mathcal{O} \left( N^{-2}\right) \right) \right] \nonumber \\
= \ & (-1)^{K-1}  N^{(K-1)(K-2)/2}    C_{\alpha,\gamma,K}^{SO(2N)}  ( \pi \mathrm{i})   \frac{ \mathrm{e}^{N\alpha} }{z\left( \alpha + \gamma\right) }    \nonumber \\
&\times  \oint _0 \dots \oint_0   \displaystyle \prod_{1 \leq j < k \leq K-1}\left[ \frac 1 {w_j + w_k}  + \frac 1 {2N}  \right] \exp\left(\displaystyle \sum_{l = 1} ^{K-1} w_l \right) \Delta \left( w_1^2, \dots , w_{K-1}^2\right)^2\nonumber \\
\times& \displaystyle \prod_{n=1}^{K-1}  \left[ \left( z(\alpha) + \frac{w_n} N z(\alpha)z(-\alpha) \right)   \frac 1 {z(\gamma)} \left( 1 - \frac{w_n} N z(-\gamma)   \right)  w_n^{3 - 2K} \mathrm{d} w_n  \right]   \left(1  + \mathcal{O}\left(N^{-2} \right)\right).
\end{align}
We now rewrite this so that all the terms take a similar form:
\begin{align} \label{eq:T_SO}
\mathcal{T}^{SO(2N)}_1 = \  (-1)^{K-1} N^{(K-1)(K-2)/2}  C_{\alpha,\gamma,K}^{SO(2N)} ( \pi \mathrm{i}) \mathrm{e}^{N\alpha}    \frac{z(\alpha)^{K-1}}{z(\alpha + \gamma) z(\gamma)^{K-1} } \left(1  + \mathcal{O}\left(N^{-2} \right)\right) \nonumber \\
 \times \left[ \oint \dots \oint \displaystyle \prod_{1 \leq j < k \leq K-1} \left(\frac{1}{w_j + w_k} \right) \exp \left( \displaystyle \sum_{l = 1}^{K-1}  w_l \right)  \Delta \left( w_1^2, \dots , w_{K-1}^2\right)^2 \displaystyle \prod_{n=1}^{K-1} w_n ^{3 - 2K}  \mathrm{d} w_n \right. \nonumber \\
 + \frac{z( - \alpha)}N \oint \dots \oint \left(\displaystyle \sum_{m = 1}^{K-1} w_m  \right)  \displaystyle \prod_{1 \leq j < k \leq K-1} \left(\frac{1}{w_j + w_k} \right) \exp \left( \displaystyle \sum_{l = 1}^{K-1}  w_l \right)  \nonumber \\
\times \Delta(w_1^2, \dots, w_{K-1}^2 ) ^2  \displaystyle \prod_{n=1}^{K-1} w_n ^{3 - 2K}  \mathrm{d} w_n \nonumber \\
- \frac{z( - \gamma)}N \oint \dots \oint \left(\displaystyle \sum_{m = 1}^{K-1} w_m  \right)  \displaystyle \prod_{1 \leq j < k \leq K-1} \left(\frac{1}{w_j + w_k} \right) \exp \left( \displaystyle \sum_{l = 1}^{K-1}  w_l \right)  \nonumber \\
\times \Delta(w_1^2, \dots, w_{K-1}^2 ) ^2  \displaystyle \prod_{n=1}^{K-1} w_n ^{3 - 2K}  \mathrm{d} w_n \nonumber \\
+  \frac{1}{2N} \oint \dots \oint \left(\displaystyle \sum_{1\leq m < n \leq K-1}  (w_m + w_n) \right)  \displaystyle \prod_{1 \leq j < k \leq K-1} \left(\frac{1}{w_j + w_k} \right) \exp \left( \displaystyle \sum_{l = 1}^{K-1}  w_l \right) \nonumber \\
\times \left.   \Delta(w_1^2, \dots, w_{K-1}^2 ) ^2  \displaystyle \prod_{n=1}^{K-1} w_n ^{3 - 2K}  \mathrm{d} w_n \right].
\end{align}
% = \  (-1)^{K-1} C_{\alpha,\gamma,K}^{O} ( \pi \mathrm{i}) \mathrm{e}^{N\alpha}    \frac{z(\alpha)^{K-1}}{z(\alpha + \gamma) z(\gamma)^{K-1} } \left(1  + \mathcal{O}\left(N^{-2} \right)\right) \nonumber \\
% \times \left[ \oint \dots \oint \displaystyle \prod_{1 \leq j < l \leq K-1} \left(\frac{1}{w_j + w_l} \right) \exp \left( \displaystyle \sum_{l = 1}^{K-1}  w_l \right)  \Delta \left( w+1^2, \dots , w_{K-1}^2\right)^2 \displaystyle \prod_{n=1}^{K-1} w_n ^{3 - 2K}  \mathrm{d} w_n \right. \nonumber \\
% + \frac{z( - \alpha)}N \oint \dots \oint \left(\displaystyle \sum_{j = 1}^{K-1} w_j  \right)  \displaystyle \prod_{1 \leq j < k \leq K-1} \left(\frac{1}{w_j + w_k} \right) \exp \left( \displaystyle \sum_{l = 1}^{K-1}  w_l \right)  \nonumber \\
%\times \Delta(w_1^2, \dots, w_{K-1}^2 ) ^2  \displaystyle \prod_{n=1}^{K-1} w_n ^{3 - 2K}  \mathrm{d} w_n \nonumber \\
%- \frac{z( - \gamma)}N \oint \dots \oint \left(\displaystyle \sum_{j = 1}^{K-1} w_j  \right)  \displaystyle \prod_{1 \leq j < k \leq K-1} \left(\frac{1}{w_j + w_k} \right) \exp \left( \displaystyle \sum_{l = 1}^{K-1}  w_l \right)  \nonumber \\
%\times \Delta(w_1^2, \dots, w_{K-1}^2 ) ^2  \displaystyle \prod_{n=1}^{K-1} w_n ^{3 - 2K}  \mathrm{d} w_n \nonumber \\
%+  \frac{K-2}{2N} \oint \dots \oint \left(\displaystyle \sum_{j = 1}^{K-1}  w_j \right)  \displaystyle \prod_{1 \leq j < k \leq K-1} \left(\frac{1}{w_j + w_k} \right) \exp \left( \displaystyle \sum_{l = 1}^{K-1}  w_l \right) \nonumber \\
%\times \left.   \Delta(w_1^2, \dots, w_{K-1}^2 ) ^2  \displaystyle \prod_{n=1}^{K-1} w_n ^{3 - 2K}  \mathrm{d} w_n \right] 
Let us label the integrals in (\ref{eq:T_SO}):
\begin{align}\label{eq:MJ_ortho}
\mathcal{M}^{SO(2N)} = \oint \dots \oint \displaystyle \prod_{1 \leq j < k \leq K-1} \left(\frac{1}{w_j + w_k} \right) \exp \left( \displaystyle \sum_{l = 1}^{K-1}  w_l \right)  \nonumber \\
\times \Delta(w_1^2, \dots, w_{K-1}^2 ) ^2  \displaystyle \prod_{n=1}^{K-1} w_n ^{3 - 2K}  \mathrm{d} w_n , \\
\mathcal{J}^{SO(2N)} = \oint \dots \oint  \left(\displaystyle \sum_{m = 1}^{K-1} w_m  \right)\displaystyle \prod_{1 \leq j < k \leq K-1} \left(\frac{1}{w_j + w_k} \right) \exp \left( \displaystyle \sum_{l = 1}^{K-1}  w_l \right) \nonumber \\
\times   \Delta(w_1^2, \dots, w_{K-1}^2 ) ^2  \displaystyle \prod_{n=1}^{K-1} w_n ^{3 - 2K}  \mathrm{d} w_n .\label{eq:MJ_ortho2}
\end{align}
Note that the final integral in (\ref{eq:T_SO}) is just  $(K-2)\mathcal{J}^{SO(2N)}$, and in fact we will see in (\ref{eq:Jvanddeterminant2}) that  $\mathcal{J}^{SO(2N)}$ and $\mathcal{M}^{SO(2N)}$ are related by 
$\mathcal{J}^{SO(2N)} = \frac{(K-1)(K-2)}{2} \mathcal{M}^{SO(2N)}$.

%%%%%%%%%%%%%%%%%%%%%%%%%%%%%%%%%%%%%%%%%%%%%%%%%%%%%%%%%%%%%%%
\subsubsection{Evaluating $\mathcal{M}^{SO(2N)}$} \label{sect:M}

In this section we will evaluate $\mathcal{M}^{SO(2N)}$ directly because we build on the method in the following section, although we will see in Section   \ref{sect:one_pole_even1} that this integral in fact shows up in the moment of the characteristic polynomial itself and so has already indirectly been evaluated in previous literature. 

Cancelling the $\frac{1}{w_j+w_l}$ factors from (\ref{eq:MJ_ortho}) with Vandermonde factors we have
\begin{eqnarray}
\mathcal{M}^{SO(2N)} &= & \oint \dots \oint \displaystyle \prod_{1 \leq j < l \leq K-1}( w_l^2-w_j^2)(w_l-w_j) \; \exp \left( \displaystyle \sum_{l = 1}^{K-1}  w_l \right)  \displaystyle \prod_{n=1}^{K-1} w_n ^{3 - 2K}  \mathrm{d} w_n\nonumber\\
&=&(K-1)! \oint \dots \oint  \left|  \begin{array}{ccccc} 
1 & w_1^2 & w_1^4 & \dots & w_1^{2K-4} \\
1 & w_2^2 & w_2^4 & \dots & w_2^{2K-4} \\
\vdots & \vdots & \vdots & \ddots & \vdots \\
1 & w_{K-1}^2 & w_{K-1}^4 & \dots & w_{K-1}^{2K-4} \end{array} \right| \nonumber \\
&&\qquad \times w_1^{3 - 2K} w_2^{4 - 2K} \dots w_{K-1}^{1-K}\exp \left( \displaystyle \sum_{l = 1}^{K-1}  w_l \right) \mathrm{d}w_1 \dots \mathrm{d}w_{K-1}, \label{eq:Mvanddeterimant}
\end{eqnarray}
where for the second line we have used the identical method to Section \ref{sect:all_zero_even} (see the similarities with (\ref{det_even}) - the difference being that we have just $K-1$ variables).

Continuing to follow the method from Section \ref{sect:all_zero_even} we write $\mathcal{M}^{SO(2N)} $ as a determinant of gamma functions. Then, in the second line below, we write the gamma functions as factorials, and then reverse the order of the columns (acquiring an overall sign from the $(K-1)(K-2)/2$ column exchanges).  We then multiply the new $j$th column by $(2j-2)!$: 
\begin{eqnarray}\label{eq:detforM}
\mathcal{M}^{SO(2N)} &= &(2\pi i)^{K-1} (K-1)! \det_{(K-1)\times(K-1)}\left(\begin{array}{ccccc} \frac{1}{\Gamma(2K-3)} &  \frac{1}{\Gamma(2K-5)} &  \frac{1}{\Gamma(2K-7)} &\cdots & \frac{1}{\Gamma(1)} \\ \frac{1}{\Gamma(2K-4)}& \frac{1}{\Gamma(2K-6)}& \frac{1}{\Gamma(2K-8)}& \cdots & \frac{1}{\Gamma(0)} \\ \vdots &\vdots &\vdots &\ddots&\vdots \\  \frac{1}{\Gamma(K-1)}&  \frac{1}{\Gamma(K-3)}&  \frac{1}{\Gamma(K-5)} &\cdots &  \frac{1}{\Gamma(3-K)}\end{array}\right)\nonumber \\
&=& (-1)^{(K-1)(K-2)/2}(2\pi i)^{K-1} (K-1)! \prod_{j=1}^{K-1} \frac{1}{(2j-2)!} \\
&&\qquad\qquad \times\det_{(K-1)\times(K-1)} \left( \begin{array}{ccccc} \frac{0!}{0!} & \cdots & \frac{(2K-8)!}{(2K-8)!} & \frac{ (2K-6)!}{(2K-6)!} & \frac{(2K-4)!}{(2K-4)!} \\  \frac{0!}{(-1)!} & \cdots & \frac{(2K-8)!}{(2K-9)!} & \frac{ (2K-6)!}{(2K-7)!} & \frac{(2K-4)!}{(2K-5)!} \\ \vdots &\ddots&\vdots &\vdots &\vdots\\  \frac{0!}{(2-K)!} & \cdots & \frac{(2K-8)!}{(K-6)!} & \frac{ (2K-6)!}{(K-4)!} & \frac{(2K-4)!}{(K-2)!} \end{array} \right) .\nonumber
\end{eqnarray}

We now use a trick from \cite{kn:alvsna20} to turn this matrix into a Vandermonde determinant.  We note that the first row above is all 1's.  The element in the $j$th position in row two is $2j-2$. Thus we can reduce the $j$th element in the second row to just $j$ by adding twice the first row to the second row (this does not change the determinant) and then pulling a factor of 2 out of the second row.  Note that this procedure works equally well for the first entry in row two, which is zero. 

\begin{eqnarray}\label{eq:Mvanddeterminant1}
\mathcal{M}^{SO(2N)} &= &2\times(-1)^{(K-1)(K-2)/2}(2\pi i)^{K-1} (K-1)! \prod_{j=1}^{K-1} \frac{1}{(2j-2)!} \\
&&\qquad \times\det_{(K-1)\times(K-1)} \left( \begin{array}{cccccc} 1&1& \cdots & 1 & 1&1 \\  1&2&\cdots& (K-3)&(K-2)&(K-1)\\\frac{0!}{(-2)!} &\frac{2!}{0!}& \cdots & \frac{(2K-8)!}{(2K-10)!} & \frac{ (2K-6)!}{(2K-8)!} & \frac{(2K-4)!}{(2K-6)!} \\ \vdots &\vdots &\ddots&\vdots &\vdots &\vdots\\  \frac{0!}{(2-K)!} & \frac{2!}{(4-K)!}&\cdots & \frac{(2K-8)!}{(K-6)!} & \frac{ (2K-6)!}{(K-4)!} & \frac{(2K-4)!}{(K-2)!} \end{array} \right) .\nonumber
\end{eqnarray}

We note that in the rest of the matrix the $j$th entry in the $m$th row is a polynomial in $j$ of order $m-1$ with leading coefficient $2^{m-1}$: for example, the $j$th entry in the third row is $(2j-2)(2j-3)$.  Working one row at a time, all but the leading order term of the polynomial can be removed by adding multiples of rows higher up the matrix, and the power of 2 is pulled out of the determinant. This leaves us with

\begin{eqnarray}
\mathcal{M}^{SO(2N)} &= &(-1)^{(K-1)(K-2)/2}(2\pi i)^{K-1} 2^{(K-1)(K-2)/2}(K-1)! \prod_{j=1}^{K-1} \frac{1}{(2j-2)!} \\
&&\quad \times\det_{(K-1)\times(K-1)} \left(\begin{array}{cccccc} 1&1&\cdots&1&1&1\\1&2&\cdots&(K-3)&(K-2)&(K-1)\\ \vdots & \vdots & \ddots & \vdots & \vdots & \vdots \\1 & 2^{K-2} &\cdots & (K-3)^{K-2}& (K-2)^{K-2} &(K-1)^{K-2} \end{array}\right) \nonumber \\
&=& (-1)^{(K-1)(K-2)/2}(2\pi i)^{K-1} 2^{(K-1)(K-2)/2}(K-1)! \prod_{j=1}^{K-1} \frac{(j-1)!}{(2j-2)!} \nonumber \\
&=& (-1)^{(K-1)(K-2)/2}(2\pi i)^{K-1} (K-1)! \prod_{j=1}^{K-2} \frac{1}{(2j-1)!!} ,
\end{eqnarray}
where we have used that the Vandermonde $V(1,2,3,\ldots,n)=\prod_{j=1}^n(j-1)!$.

%%%%%%%%%%%%%%%%%%%%%%%%%%%%%%%%%%%%%%%%%%%%%%%%%%%%%%%%%%%%%%%
\subsubsection{Evaluating $\mathcal{J}^{SO(2N)}$} \label{sect:J}

As in (\ref{eq:Mvanddeterimant}), we have
\begin{eqnarray}
\mathcal{J}^{SO(2N)} &=&(K-1)! \oint \dots \oint  \left|  \begin{array}{ccccc} 
1 & w_1^2 & w_1^4 & \dots & w_1^{2K-4} \\
1 & w_2^2 & w_2^4 & \dots & w_2^{2K-4} \\
\vdots & \vdots & \vdots & \ddots & \vdots \\
1 & w_{K-1}^2 & w_{K-1}^4 & \dots & w_{K-1}^{2K-4} \end{array} \right| \left(\sum_{m=1}^{K-1} w_m\right) \nonumber \\
&&\qquad \times w_1^{3 - 2K} w_2^{4 - 2K} \dots w_{K-1}^{1-K}\exp \left( \displaystyle \sum_{l = 1}^{K-1}  w_l \right) \mathrm{d}w_1 \dots \mathrm{d}w_{K-1}.\label{eq:Jvanddeterimant}
\end{eqnarray}

Using the usual method from Section \ref{sect:all_zero_even} (noting that the integrand of (\ref{eq:Jvanddeterimant}) is symmetric as long as we retain the sum over $m$ intact) to pull all the factors containing $w_j$ into the $j$th row of the determinant, and also bringing the integration inside the determinant and evaluating the matrix elements as gamma functions, we see that each term in the sum over $m$ gives zero except for the $m=K-1$ term.  This is because all terms except the $K-1$th raise the power on $w_m$, $m=1,\ldots,K-2$ by one, making the power equal to that on $w_{j+1}$.  This results in two identical rows of gamma functions, and so the determinant is zero.  For $m=K-1$, however, we get the following non-zero determinant, which has a jump of 2 in the argument of the gamma function between the second-to-last and last row instead of the usual change by 1:

\begin{eqnarray}
\mathcal{J}^{SO(2N)} &= &(2\pi i)^{K-1} (K-1)! \det_{(K-1)\times(K-1)}\left(\begin{array}{ccccc} \frac{1}{\Gamma(2K-3)} &  \frac{1}{\Gamma(2K-5)} &  \frac{1}{\Gamma(2K-7)} &\cdots & \frac{1}{\Gamma(1)} \\ \frac{1}{\Gamma(2K-4)}& \frac{1}{\Gamma(2K-6)}& \frac{1}{\Gamma(2K-8)}& \cdots & \frac{1}{\Gamma(0)} \\ \vdots &\vdots &\vdots &\ddots&\vdots\\  \frac{1}{\Gamma(K)}&  \frac{1}{\Gamma(K-2)}&  \frac{1}{\Gamma(K-4)} &\cdots &  \frac{1}{\Gamma(4-K)} \\  \frac{1}{\Gamma(K-2)}&  \frac{1}{\Gamma(K-4)}&  \frac{1}{\Gamma(K-6)} &\cdots &  \frac{1}{\Gamma(2-K)}\end{array}\right) \nonumber \\
&=& (-1)^{(K-1)(K-2)/2}(2\pi i)^{K-1} (K-1)! \prod_{j=1}^{K-1} \frac{1}{(2j-2)!} \\
&&\qquad \times\det_{(K-1)\times(K-1)} \left( \begin{array}{cccccc} \frac{0!}{0!} &\frac{2!}{2!}& \cdots & \frac{(2K-8)!}{(2K-8)!} & \frac{ (2K-6)!}{(2K-6)!} & \frac{(2K-4)!}{(2K-4)!} \\  \frac{0!}{(-1)!} &\frac{2!}{1!}& \cdots & \frac{(2K-8)!}{(2K-9)!} & \frac{ (2K-6)!}{(2K-7)!} & \frac{(2K-4)!}{(2K-5)!} \\\vdots & \vdots &\ddots&\vdots &\vdots &\vdots\\  \frac{0!}{(3-K)!} & \frac{2!}{(5-K)!}&\cdots & \frac{(2K-8)!}{(K-5)!} & \frac{ (2K-6)!}{(K-3)!} & \frac{(2K-4)!}{(K-1)!} \\  \frac{0!}{(1-K)!} & \frac{2!}{(3-K)!}& \cdots & \frac{(2K-8)!}{(K-7)!} & \frac{ (2K-6)!}{(K-5)!} & \frac{(2K-4)!}{(K-3)!} \end{array} \right) .\nonumber
\end{eqnarray}

As in the previous section, we note that for the first $K-2$ rows the matrix element in the $j$th column and $m$th row is a polynomial in $j$ of order $m-1$.  For example, the $j$th element in the 2nd row is just $2j-2$ and the $j$th element in the $(K-2)$th row is $(2j-2)(2j-3)\cdots(2j-(K-2))$.  Working sequentially from the top, each polynomial can be reduced to just its highest order term, $(2j)^{m-1}$ by adding multiples of previous rows.  The final row, the only one that differs from (\ref{eq:Mvanddeterminant1}), contains polynomials of order $K-1$: $(2j-2)(2j-3)\ldots(2j-K)=(2j)^{K-1}-(2j)^{K-2}(2+3+\cdots+K) +\mathcal{O}(j^{K-3})$. So, splitting the determinant into two determinants, each one containing just one of the two surviving terms from the bottom row, we have

\begin{eqnarray}\label{eq:Jvanddeterminant1}
&&\mathcal{J}^{SO(2N)} = (-1)^{(K-1)(K-2)/2}(2\pi i)^{K-1} 2^{(K-1)(K-2)/2}(K-1)! \prod_{j=1}^{K-1} \frac{1}{(2j-2)!} \\
&&\qquad \times\left[2\det_{(K-1)\times(K-1)} \left( \begin{array}{cccccc} 1&1& \cdots &1& 1 & 1 \\ 1&2& \cdots & (K-3) & (K-2)& (K-1) \\\vdots & \vdots &\ddots&\vdots &\vdots &\vdots\\ 1^{K-3}& 2^{K-3}&\cdots & (K-3)^{K-3} & (K-2)^{K-3}& (K-1)^{K-3}\\ 1^{K-1} &2^{K-1}& \cdots &(K-3)^{K-1}& (K-2)^{K-1} & (K-1)^{K-1}\end{array} \right) \right.\nonumber\\&&\nonumber\\
&&\qquad -  (2+3+\cdots+K)\nonumber \\
&&\qquad\;\;\times\left.\det_{(K-1)\times(K-1)} \left( \begin{array}{cccccc} 1&1& \cdots &1& 1 & 1 \\ 1&2& \cdots & (K-3) & (K-2)& (K-1) \\\vdots & \vdots &\ddots&\vdots &\vdots &\vdots\\ 1^{K-3}& 2^{K-3}&\cdots & (K-3)^{K-3} & (K-2)^{K-3}& (K-1)^{K-3}\\ 1^{K-2} &2^{K-2}& \cdots &(K-3)^{K-2}& (K-2)^{K-2} & (K-1)^{K-2}\end{array} \right) \right]\nonumber.
\end{eqnarray}
The second determinant above is the Vandermonde $\Delta(1,2,3, \ldots K-1)= \prod_{j=1}^{K-1}(j-1)!$, while the first determinant is what is called in \cite{kn:heineman29} a ``Vandemondian" (the notation for this matrix in that paper is $V_{K-1\;\;1}$).  Theorem I of \cite{kn:heineman29} tells us that
\begin{equation}
\left| \begin{array}{ccccc}1&1&1&\cdots &1\\x_1&x_2&x_3&\cdots &x_n\\x_1^2&x_2^2&x_3^2&\cdots & x_n^2\\ \vdots & \vdots & \vdots & \ddots & \vdots \\ x_1^{n-2}&x_2^{n-2}&x_3^{n-2} &\cdots &x_n^{n-2} \\x_1^n&x_2^n&x_3^n & \cdots &x_n^n \end{array} \right| = (x_1+x_2+\cdots+x_n)\left| \begin{array}{ccccc}1&1&1&\cdots &1\\x_1&x_2&x_3&\cdots &x_n\\x_1^2&x_2^2&x_3^2&\cdots & x_n^2\\ \vdots & \vdots & \vdots & \ddots & \vdots \\ x_1^{n-2}&x_2^{n-2}&x_3^{n-2} &\cdots &x_n^{n-2} \\x_1^{n-1}&x_2^{n-1}&x_3^{n-1} & \cdots& x_n^{n-1} \end{array} \right| 
\end{equation}
This gives us a relation between the two determinants in (\ref{eq:Jvanddeterminant1}) and so we have
\begin{eqnarray}\label{eq:Jvanddeterminant2}
\mathcal{J}^{SO(2N)} &=&(-1)^{(K-1)(K-2)/2}(2\pi i)^{K-1} 2^{(K-1)(K-2)/2}(K-1)! \prod_{j=1}^{K-1} \frac{1}{(2j-2)!}\nonumber\\
\nonumber &&\qquad\times \frac{(K-1)(K-2)}{2} \Delta(1,2,3,\ldots,K-1)\\
&=& \frac{(K-1)(K-2)}{2} \mathcal{M}^{SO(2N)}.
\end{eqnarray}

Note we have inadvertently proven the interesting determinant relation:
\begin{eqnarray}
&&\frac{(K-1)(K-2)}{2}\det_{(K-1)\times(K-1)}\left(\begin{array}{ccccc} \frac{1}{\Gamma(2K-3)} &  \frac{1}{\Gamma(2K-5)} &  \frac{1}{\Gamma(2K-7)} &\cdots & \frac{1}{\Gamma(1)} \\ \frac{1}{\Gamma(2K-4)}& \frac{1}{\Gamma(2K-6)}& \frac{1}{\Gamma(2K-8)}& \cdots & \frac{1}{\Gamma(0)} \\ \vdots &\vdots &\vdots &\ddots&\vdots \\ \frac{1}{\Gamma(K)}&  \frac{1}{\Gamma(K-2)}&  \frac{1}{\Gamma(K-4)} &\cdots &  \frac{1}{\Gamma(4-K)} \\  \frac{1}{\Gamma(K-1)}&  \frac{1}{\Gamma(K-3)}&  \frac{1}{\Gamma(K-5)} &\cdots &  \frac{1}{\Gamma(3-K)}\end{array}\right)\\
&&\qquad \qquad \qquad = \det_{(K-1)\times(K-1)}\left(\begin{array}{ccccc} \frac{1}{\Gamma(2K-3)} &  \frac{1}{\Gamma(2K-5)} &  \frac{1}{\Gamma(2K-7)} &\cdots & \frac{1}{\Gamma(1)} \\ \frac{1}{\Gamma(2K-4)}& \frac{1}{\Gamma(2K-6)}& \frac{1}{\Gamma(2K-8)}& \cdots & \frac{1}{\Gamma(0)} \\ \vdots &\vdots &\vdots &\ddots&\vdots\\  \frac{1}{\Gamma(K)}&  \frac{1}{\Gamma(K-2)}&  \frac{1}{\Gamma(K-4)} &\cdots &  \frac{1}{\Gamma(4-K)} \\  \frac{1}{\Gamma(K-2)}&  \frac{1}{\Gamma(K-4)}&  \frac{1}{\Gamma(K-6)} &\cdots &  \frac{1}{\Gamma(2-K)}\end{array}\right).\nonumber
\end{eqnarray}

We have already seen  that the third integral in (\ref{eq:T_SO}) is 
\begin{align}\label{eq:J2ortho}
 \oint \dots \oint  \left(\displaystyle \sum_{1 \leq m <n \leq K-1} \left(  w_m + w_n \right) \right)\displaystyle \prod_{1 \leq j < l \leq K-1} \left(\frac{1}{w_j + w_l} \right) \exp \left( \displaystyle \sum_{k = 1}^{K-1}  w_k \right) \nonumber \\
\times   \Delta(w_1^2, \dots, w_{K-1}^2 ) ^2  \displaystyle \prod_{n=1}^{K-1} w_n ^{3 - 2K}  \mathrm{d} w_n \nonumber \\
=  (K-2)\mathcal{J}^{SO(2N)}.
\end{align}

%%%%%%%%%%%%%%%%%%%%%%%%%%%%%%%%%%%%%%%%%%%%%%%%%%%
\subsubsection{Pulling together the case of one pole away from zero} \label{sect:one_pole_even1}
We now feed the results of Sections \ref{sect:M} and \ref{sect:J} into  (\ref{eq:T_SO}), giving
\begin{align}\label{ntlot_ortho}
\mathcal{T}^{SO(2N)}_1 =   (-1)^{K-1}  C_{\alpha,\gamma,K}^{SO(2N)}N^{(K-1)(K-2)/2}   ( \pi \mathrm{i}) \mathrm{e}^{N\alpha}    \frac{z(\alpha)^{K-1}}{z(\alpha + \gamma) z(\gamma)^{K-1} } \left(1  + \mathcal{O}\left(N^{-2} \right)\right) \nonumber \\
 \times \left[ \mathcal{M}^{SO(2N)} +  \frac{\mathcal{J}^{SO(2N) }}{N} \left( z(-\alpha) - z(-\gamma)   + \frac{K-2}{2}  \right)  \right]\nonumber\\
 = (-1)^{K-1} C_{\alpha,\gamma,K}^{SO(2N)} N^{(K-1)(K-2)/2} ( \pi \mathrm{i})\mathrm{e}^{N\alpha}    \frac{z(\alpha)^{K-1}}{z(\alpha + \gamma) z(\gamma)^{K-1} } \left(1  + \mathcal{O}\left(N^{-2} \right)\right) \nonumber \\
\times \ \mathcal{M}^{SO(2N)} \left(1+ \frac {(K-1)(K-2)} {2N} \left(z (-\alpha) - z(-\gamma) + \frac{K-2}{2} \right) \right),
\end{align}
where in the second equality we have used (\ref{eq:Jvanddeterminant2}). 

The corresponding contribution from a pole at $-\alpha$ is simply
\begin{align}
\mathcal{T}^{SO(2N)}_2  =  (-1)^{K-1} C_{\alpha,\gamma,K}^{SO(2N)} N^{(K-1)(K-2)/2} (\pi \mathrm{i}) \mathrm{e}^{-N\alpha}    \frac{z(-\alpha)^{K-1}}{z(-\alpha + \gamma) z(\gamma)^{K-1} } \left(1  + \mathcal{O}\left(N^{-2} \right)\right) \nonumber \\
\times \mathcal{M}^{SO(2N)} \left(1  + \frac {(K-1)(K-2) } {2N} \left(z (\alpha) - z(-\gamma) + \frac{K-2}{2} \right) \right).
\end{align}
The total contribution from all the terms with one pole evaluated away from zero is
\begin{align}
&K \mathcal{T}^{SO(2N)}_1 + K \mathcal{T}^{SO(2N)}_2 = \nonumber \\
& \frac{ K (-1)^{K-1} C_{\alpha,\gamma,K}^{SO(2N)} N^{(K-1)(K-2)/2} ( \pi \mathrm{i})  \mathcal{M}^{SO(2N)}  }{z(\gamma)^{K-1}} \left(1  + \mathcal{O}\left(N^{-2} \right)\right)\nonumber \\
&\times  \left[ \mathrm{e}^{N\alpha}    \frac{z(\alpha)^{K-1}}{z(\alpha + \gamma)  }\left(1+ \frac {(K-1)(K-2)} {2N} \left(z (-\alpha) - z(-\gamma) + \frac{K-2}{2} \right) \right)   \right. \nonumber \\
& + \left. \mathrm{e}^{-N\alpha}    \frac{z(-\alpha)^{K-1}}{z(-\alpha + \gamma)  }\left(1+ \frac {(K-1)(K-2)} {2N} \left(z (\alpha) - z(-\gamma) + \frac{K-2}{2} \right) \right)  \right].
\end{align}

We note that another approach would be to relate the quantities  $\mathcal{M}^{SO(2N)}$ and  $\mathcal{J}^{SO(2N)}$  to the leading order terms of the well-studied moment of characteristic polynomials obtained by setting $\alpha=\gamma$.  The moment for finite matrix size $N$ can also be written as a multiple contour integral \cite{kn:cfkrs}
\begin{align}\label{orth_mom}
&\int _{A \in SO(2N)} \Lambda(1)^{K-1} \mathrm{d}A = 
(-1)^{(K-1)(K-2)/2} 2^{K-1} (2\pi \mathrm{i})^{-K+1}((K-1)!)^{-1} \nonumber  \\
&\times \oint \dots \oint \frac{ \displaystyle \prod _{1 \leq j < k \leq K-1} z \left(w_j + w_k\right) \exp\left(N \sum_{l = 1} ^{K-1} w_l\right) \Delta (w_1^2, w_2^2, \dots, w_{K-1}^2)^2 \displaystyle \prod _{i = 1} ^{K-1} w_i \mathrm{d}w_i     }{\displaystyle \prod _{n = 1} ^{K-1} w_n ^{2K-2} }  ,
\end{align}
where the contours of integration encircle zero.

Upon scaling the $w$ variables by $1/N$ in the manner of Section \ref{sect:all_zero_even} we find the relation with the integrals $\mathcal{M}^{SO(2N)}$ and  $\mathcal{J}^{SO(2N)}$ from (\ref{eq:MJ_ortho}).
\begin{eqnarray}\label{eq:momentMJ}
\int _{A \in SO(2N)} \Lambda(1)^{K-1} \mathrm{d}A &= & (-1)^{(K-1)(K-2)/2} 2^{K-1} (2\pi \mathrm{i})^{-K+1} ((K-1)!)^{-1} \\
&&\qquad \times N^{(K-1)(K-2)/2 }   \left( \mathcal{M}^{SO(2N)}    + \frac{K-2}{2N} \mathcal{J}^{SO(2N)}+\mathcal{O}\left(\frac {1}{ N^2} \right)\right). \nonumber
\end{eqnarray}
The moment was calculated in \cite{kn:keasna00b} using the Selberg integral  and the leading order was expressed as (also obtained independently by  \cite{kn:brezhik00})
\begin{align} \label{values_formula}
\int _{A \in SO(2N)} \Lambda(1)^{K-1} \mathrm{d}A& = N^{(K-1)(K-2)/2} 2 ^{(K-1)^2/2} \frac{G(K) \sqrt{\Gamma(2K-1) }} {\sqrt{G(2K-1) \Gamma(K)}} \left( 1 + \mathcal{O}\left(\frac 1 N \right)\right)\nonumber \\
&=N^{(K-1)(K-2)/2} 2^{K-1}\prod_{j=1}^{K-2} \frac{1}{(2j-1)!!}\left( 1 + \mathcal{O}\left(\frac 1 N \right)\right),
\end{align}
where $G$ is the Barnes $G$-function introduced in \cite{kn:barnes00}. (The $G$-function is uniquely defined by the functional equation
\begin{equation}
G(s+1) = \Gamma(s) G(s)
\end{equation}
and the conditions that $G(0) = 0$, $\frac{ \text{d}^3 }{\text{d} s^3} G(s) \geq 0$.)  So we see that we have a representation of $\mathcal{M}^{SO(2N)}$ in terms of Barnes $G$-functions:

\begin{align}\label{eq:barnesGforM}
 \mathcal{M}^{SO(2N)}  = (-1)^{(K-1)(K-2)/2}  (2\pi \mathrm{i})^{K-1}  2 ^{(K-1)(K-3)/2} \frac{G(K) \sqrt{\Gamma(2K-1)  \Gamma(K)}} {\sqrt{G(2K-1)}}   .
\end{align}

%%%%%%%%%%%%%%%%%%%%%%%%%%%%%%%%%%%%%%%%%%%%%%%%%
\subsection{Completing the mixed moment for $SO(2N)$}

Returning to  (\ref{sum_even_ortho}), we have found that in the sum over the $\epsilon$'s the leading and next-to-leading order terms when $N$ is large result from one variable being integrated around a pole at $\alpha$ or $-\alpha$ with all the other variables integrated around the pole at zero.  The leading order term is of order $N^{(K-1)(K-2)/2}$ and when all poles are evaluated around zero the result is at most order $N^{\tfrac{1}{2}K^2-\tfrac{5}{2}K-1}$ so this is at least two orders lower than the leading order as long as $K\geq 0$.   

The leading order and next-to-leading order terms were evaluated in the previous section. Also recall that
\begin{equation} 
C^{SO(2N)}_{\alpha, \gamma, K} = \frac{ \mathrm{e}^{-N \alpha } (-1)^{K(K-1)/2} 2^K z(2 \gamma) }{(2 \pi \mathrm{i})^K  \Gamma(K+1)}.
\end{equation}

 Plugging these into (\ref{sum_even_ortho}) and replacing $K-1$ with $r$, we have
\begin{align}\label{eq:even_moment_r}
& \int _{SO(2N)}\frac{\Lambda(1)^{r}  {\Lambda ( \mathrm{e}^{-\alpha} )}}{\Lambda ( \mathrm{e}^{- \gamma} )}  \mathrm{dA} \nonumber \\
 =\ &   2^{r^2/2 }    N^{r(r-1)/2}   \left(1  + \mathcal{O}\left(N^{-2} \right)\right) \nonumber \\
&\times   \left[ \frac{ z(2 \gamma)  z(\alpha)^{r}}{z(\alpha + \gamma) z(\gamma)^{r} }   \left(1  + \frac {r(r-1) } {2N} \left(z (-\alpha) - z(-\gamma) + \frac{r-1}{2} \right) \right)  \right. \nonumber \\
&+ \frac{\mathrm{e}^{-2 N \alpha }  z(2 \gamma)  z(-\alpha)^{r}}{z(-\alpha + \gamma) z(\gamma)^{r} }
 \left.  \left(1  + \frac {r(r-1) } {2N} \left(z (\alpha) - z(-\gamma) + \frac{r-1}{2} \right) \right) \right] \nonumber \\
& \times   \frac{G(r+1) \sqrt{\Gamma(2r+1)}}{\sqrt{G(2r+1) \Gamma(r+1)}}.
\end{align}

 To get the logarithmic derivative, we first differentiate with respect to $\alpha$, noting that $z'(x) = z(x) z(-x)$, and that $\mathrm{d}/ \mathrm{d}x \left[ z(x)^{-1}\right]  = \mathrm{e}^{-x}$. Then, we set $\alpha = \gamma = \phi$, where ${\rm Re}(\phi)>0$:
%\begin{align}
%\int _{SO(2N)} - \mathrm{e}^{-\alpha}  \frac{\Lambda(1)^{r}  {\Lambda' ( \mathrm{e}^{-\alpha} )}}{\Lambda ( \mathrm{e}^{- \gamma} )}  \mathrm{dA} \nonumber \\
%=    2^{r^2/2}  N^{r(r-1)/2}  \frac{ z(2 \gamma) }{z(\gamma)^r}   \left(1  + \mathcal{O}\left(N^{-2} \right)\right) \nonumber \\ 
%\times \left[ \left(\frac{ r z(\alpha)^r z(-\alpha) }{z(\alpha + \gamma) } + z(\alpha)^r \mathrm{e}^{- \alpha - \gamma}  \right) \left(1 + \frac{r(r-1)}{2N} \left( z(-\alpha) - z( -\gamma) + \frac{r-1}{2} \right)    \right)  \right.  \nonumber \\
%- \frac{z(\alpha)^{r+1} z(-\alpha) }{z(\alpha + \gamma)} \frac{r(r-1)}{2N} \nonumber \\
%- \mathrm{e}^{-2N \alpha} \left(  \frac{r z(-\alpha)^r z(\alpha)}{z(-\alpha + \gamma)}   + z(- \alpha)^r  \mathrm{e}^{\alpha - \gamma} \right) \left(1 + \frac{r(r-1)}{2N} \left( z(\alpha) - z(-\gamma) + \frac{r-1}{2}\right)  \right.  \nonumber \\
%+ \left. \left. z(- \alpha)^{r+1} z(\alpha)   \frac{r(r-1)}{2N} \right)   \right.\nonumber \\
%- 2N \frac{\mathrm{e}^{-2 N \alpha }   z(-\alpha)^{r}}{z(-\alpha + \gamma) }
% \left.  \left(1  + \frac {r(r-1) } {2N} \left(z (\alpha) - z(-\gamma) + \frac{r-1}{2} \right) \right) \right] \nonumber \\
%\times   \frac{G(r+1) \sqrt{\Gamma(2r+1)}}{\sqrt{G(2r+1) \Gamma(r+1)}},
%\end{align}
\begin{eqnarray} \label{mixed_moment_ortho}
&&\int _{SO(2N)} -  \mathrm{e}^{- \phi} \frac{\Lambda(1)^{r}  {\Lambda' ( \mathrm{e}^{-\phi} )}}{\Lambda ( \mathrm{e}^{- \phi} )}  \mathrm{dA}  \\
&&=  2^{r^2/2}  N^{r(r-1)/2}   \left(1  + \mathcal{O}\left(N^{-2} \right)\right)\times   \frac{G(r+1) \sqrt{\Gamma(2r+1)}}{\sqrt{G(2r+1) \Gamma(r+1)}}  \nonumber \\ 
&&\times \left[  \left( r z(- \phi) - z(-2\phi) \right)\left(1 + \frac{r(r-1)^2}{2N}    \right) - z(\phi) z(-\phi) \frac{r(r-1)}{2N}  \right. \nonumber \\
&& \left. - \mathrm{e}^{- 2N \phi} \left(  z(2 \phi) \frac{ z(-\phi)^r}{z(\phi)^r} \left( 1 + \frac{r(r-1)}{2N} \left( z(\phi) - z(- \phi) + \frac{r-1}{2}  \right)  \right)  \right) \right] ,\nonumber
\end{eqnarray}
where we have used $z(2 \phi )\mathrm{e}^{-2 \phi} = -z(-2\phi)$. 

%%%%%%%%%%%%%%%%%%%%%%%%%%%%%%%%%%%%%%%%%%%%%%
\section{A mixed moment for $USp(2N)$}\label{sect:mixed_moment_symp}

Note that the method in Section \ref{sect:mixed_moment_even} may also be used to find the equivalent quantity over the symplectic group. To do so, one starts from the contour integral statement of the ratios theorem (from Theorem 4.2 and Lemma 6.8 in \cite{kn:cfz2}):
\begin{align}\label{ratios_mess_symp}
&\mathcal{R}_{USp(2N)}(\alpha,\gamma,K) := \int _{USp(2N)} \frac{\Lambda(1)^{K-1} \Lambda(\mathrm{e}^{-\alpha}) }{\Lambda ( \mathrm{e}^{-\gamma} )}  \mathrm{dA} = \frac{\mathrm{e}^{-N \alpha } (-1)^{K(K-1)/2} 2^K}{(2 \pi \mathrm{i})^K K!} \nonumber \\
&\times \oint \frac{  \displaystyle \prod_{1 \leq j \leq k \leq K} z (w_j + w_k) \exp \left( N\displaystyle \sum_{l=1} ^K w_l \right) \Delta ( w_1^2, \dots , w_K^2)^2 \prod_{i = 1} ^K w_i \mathrm{d}w_i  } { \displaystyle \prod_{n = 1} ^ K z(w_n + \gamma)  (w_n - \alpha)(w_n + \alpha)   w_n ^ {2K-2}  },
\end{align}
where $z(x) = (1 - \mathrm{e}^{-x})^{-1} = \left(  \frac1x + \frac1{2} + \mathcal{O}\left(x\right)    \right)$, ${\rm Re}(\gamma)>0$, and the contours of integration enclose 0, $ \alpha$ and $-\alpha$. Following the steps in Section  \ref{sect:mixed_moment_even} exactly, one finds 
 \begin{eqnarray} \label{mixed_moment_symp}
&&\int _{USp(2N)} -  \mathrm{e}^{- \phi} \frac{\Lambda(1)^{r}  {\Lambda' ( \mathrm{e}^{-\phi} )}}{\Lambda ( \mathrm{e}^{- \phi} )}  \mathrm{dA} \\
&&=  2^{r(r-2)/2}  N^{r(r+1)/2}   \left(1  + \mathcal{O}\left(N^{-2} \right)\right)   \times   \frac{G(r+1) \sqrt{\Gamma(r+1)}}{\sqrt{G(2r+1) \Gamma(2r+1)}}\nonumber \\ 
&&\times \left[ \left( 2 z(2 \phi)z(-2\phi) +    r z(- \phi) - z(-2\phi) \right)  \left(1 + \frac{r(r-1)(r+1)}{2N}    \right) \right.  \nonumber \\
&&\left.- z(2\phi) z(\phi) z(-\phi) \frac{r(r+1)}{2N}  \right. \nonumber \\
&&+ \left. \mathrm{e}^{- 2N \phi}   z(2 \phi) z(-2 \phi) \frac{ z(-\phi)^r}{z(\phi)^r} \left( 1 + \frac{r(r+1)}{2N} \left( z(\phi) - z(- \phi) + \frac{r+1}{2}  \right)  \right)    \right]. \nonumber 
\end{eqnarray}

%%%%%%%%%%%%%%%%%%%%%%%%%%%%%%%%%%%%%%%%

\section{Analytic continuation in $r$}
Until this point, $r$ has been constrained to be an integer greater than zero. However, have now written our moment as an expression which permits analytic continuation in $r$. We might hope that we can now relax the constraint that $r$ must be an integer and analytically continue $r$ in the half plane $\mathbb{R}\text{e}[r] > 0$. As can be seen in Table \ref{tab:mixedmoms}, this may well be the case (note that we would not expect perfect agreement for finite $N$ since (\ref{mixed_moment_ortho}) only captures the first two terms in the large $N$ expansion).  Given that there is some evidence that the analytic continuation matches the ensemble average, we will proceed with the one-level density application.
\begin{table}[]
\centering
\caption{Numerically generated values of $\int _{SO(200)} -  \mathrm{e}^{- \phi} \frac{\Lambda(1)^{r}  {\Lambda' ( \mathrm{e}^{-\phi} )}}{\Lambda ( \mathrm{e}^{- \phi} )}  \mathrm{dA}$, compared with the prediction of (\ref{mixed_moment_ortho}). We have chosen $\phi = 2.0 + 3.5$i for this test. The numerical values were calculated using $10^6$ matrices generated uniformly with respect to Haar measure on $SO(200)$.}
\label{tab:mixedmoms}
\begin{tabular}{lll}
$r$       & Predicted value      & Numerical value     \\ \hline
1.0       & 0.255807-0.0993974i  & 0.26529-0.103092i   \\
2.0       & 97.3408-35.0436i     & 98.0627-36.3916i    \\
0.5     & -0.031109-0.0273784i & -0.04091-0.02978i   \\
1.0+1.0i   & -0.002565+0.007075i  & -0.003478+0.009805i \\
0.5+1.0i & 0.049486-0.021654i   & 0.052749-0.023028i 
\end{tabular}
\end{table}

\section{An application of mixed moments - calculating the one level density of eigenvalues for the excised model} 
 \label{sect:old_even}

The excised ensemble $T_{\chi}\mathrm(2N)$ is the ensemble of matrices in $SO(2N)$, introduced in \cite{kn:dhkms12}, whose characteristic polynomial $\Lambda(s)$ evaluated at $1$ is greater than a cut-off value $e^{\chi}$.

We will denote the joint probability distribution function for the eigenvalues over $SO(2N)$ by $P(\theta_1,\dots,\theta_N)$, so that Haar measure for $SO(2N)$ can be written as: 
\begin{align}
\mathrm{dA} =& P(\theta_1,\dots,\theta_N)  \prod _{l=1} ^{N} \mathrm{d} \theta_l  \nonumber \\
=& \nonumber  S_N \prod _{1\leq j < k \leq N} \left( \cos \theta_k - \cos \theta_j \right)^2 \prod _{l=1} ^{N} \mathrm{d} \theta_l . 
\end{align}
Here, $S_N := \frac{2^{(N-1)^2}}{\pi^N N!} $ is a normalisation constant and the $\theta _l$s  are the eigenangles.

The functional equation for the characteristic polynomial of an even orthogonal matrix is
\begin{equation}
\Lambda (s) = s^{2N} \Lambda(s^{-1}) .
\end{equation}
Differentiating this gives the following identity:
\begin{equation}
s \frac{\Lambda ' (s)}{\Lambda (s)} = 2N -  s^{-1}\frac{\Lambda ' (s^{-1})}{\Lambda (s^{-1})}  .
\end{equation}
Changing variables gives
\begin{align}
\label{eq:func}
\mathrm{e}^{\mathrm{i} \phi  }\frac{  \Lambda'(\mathrm{e}^{\mathrm{i} \phi})}{ \Lambda(\mathrm{e}^{\mathrm{i} \phi})}= 2N -  \mathrm{e}^{-\mathrm{i} \phi  }\frac{  \Lambda'(\mathrm{e}^{-\mathrm{i} \phi})}{ \Lambda(\mathrm{e}^{-\mathrm{i} \phi})}.
\end{align}

Using residue calculus, we can write the Heaviside function $H(f(x) - \chi)$ as:
\begin{equation}
\label{eq:H}
H(f(x)-\chi) = \frac{1}{2 \pi \mathrm{i}} \displaystyle \int _{d-i  \infty} ^{d+i \infty} \frac{\mathrm{exp}({rf(x) - r \chi})}{r} \, \mathrm{d}r
\end{equation}
for $d > 0$.

Let \begin{align}
\mathcal{V}(N,r) := N^{r(r-1)/2} 2 ^{r^2/2} \frac{G(r+1) \sqrt{\Gamma(2r+1) }} {\sqrt{G(2r+1)   \Gamma(r+1)}}
\end{align}
and
\begin{align}
\mathcal{U}(N,r,\phi) :=  \left[  \left( r z(- \phi) - z(-2\phi) \right)\left(1 + \frac{r(r-1)^2}{2N}    \right) - z(\phi) z(-\phi) \frac{r(r-1)}{2N}  \right. \nonumber \\
\left. - \mathrm{e}^{- 2N \phi}   z(2 \phi) \frac{ z(-\phi)^r}{z(\phi)^r} \left( 1 + \frac{r(r-1)}{2N} \left( z(\phi) - z(- \phi) +  \frac{r-1}{2}   \right)   \right)\right].
\end{align}
  For the $r$th moment, we know from (\ref{eq:momentMJ}) that
\begin{eqnarray} \label{moment2}
&&\int _{SO(2N)  }\Lambda(1)^{r}  \mathrm{dA} \nonumber \\
&&\qquad=     N^{r(r-1)/2} 2 ^{r^2/2} \frac{G(r+1) \sqrt{\Gamma(2r+1) }} {\sqrt{G(2r+1)   \Gamma(r+1)}} \left( 1 + \frac{r (r-1)^2}{2N} + \mathcal{O}\left( N^{-2}\right) \right)   \nonumber \\
&&\qquad= \mathcal{V}(N,r) \left( 1 + \frac{r (r-1)^2}{2N}+  \mathcal{O}\left( N^{-2}\right) \right).
\end{eqnarray}
Using (\ref{mixed_moment_ortho}), we also have
\begin{eqnarray} \label{moment1}
&&\int _{SO(2N)}    \    - \mathrm{e}^{ - \phi}  \frac{\Lambda(1)^{r}  {\Lambda' ( \mathrm{e}^{ - \phi} )}}{\Lambda ( \mathrm{e}^{- \phi} )}  \mathrm{dA}  \nonumber \\
&&\qquad=   \mathcal{U}(N,r,\phi)    \mathcal{V}(N,r)\left (1+    + \mathcal{O}\left(N^{-2} \right)\right). 
\end{eqnarray}

\begin{figure}[h!]
\centering
  \includegraphics[width=10cm]{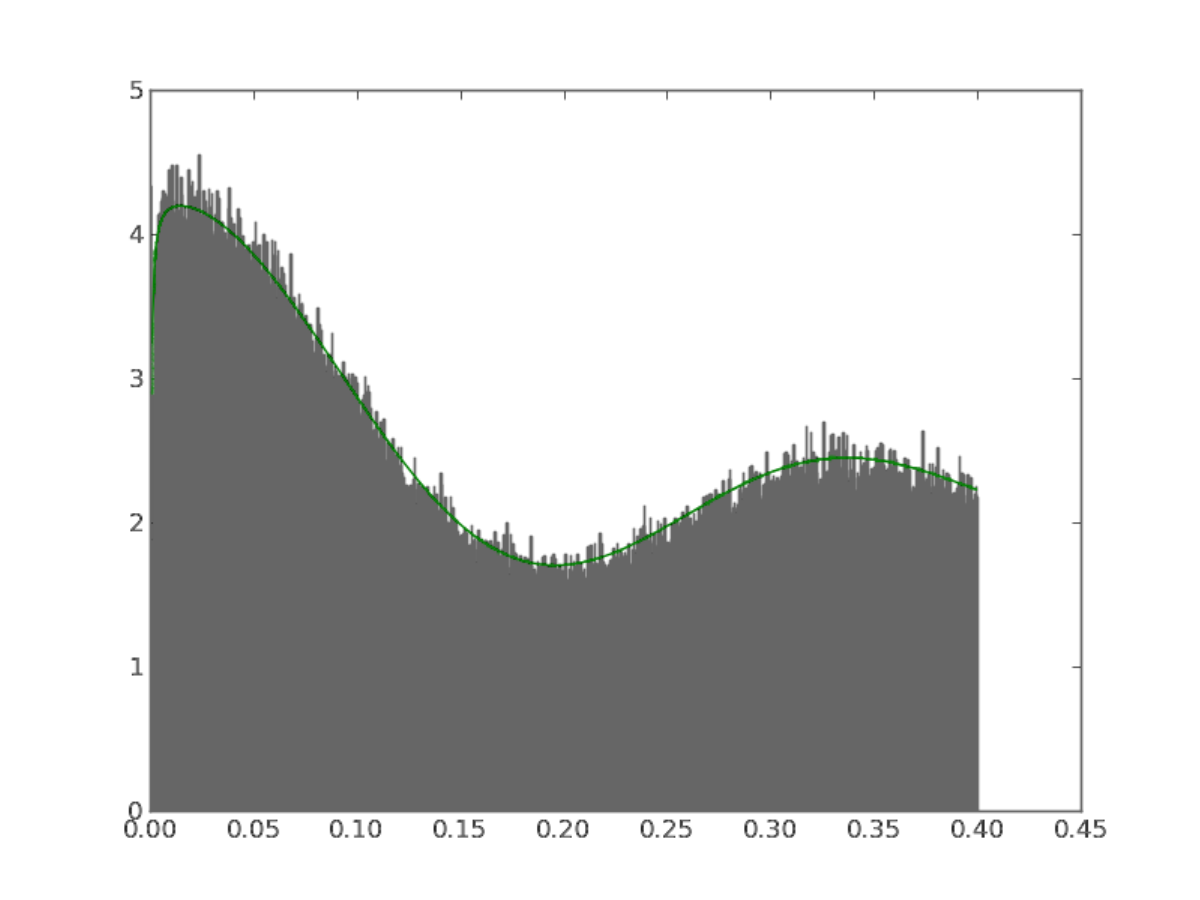}
  \caption{One level density for the excised $SO(24)$ ensemble with $\chi = \log(0.0001)$, generated numerically using $10^7$ matrices (histogram), with the first four terms (up to $k = 3$) from (\ref{residues}) plotted as the smooth curve. Both plots are normalised to have unit area under the curve.}
\label{one_level_density_ntlo}
\end{figure}

We wish to find the one level density $R_{f}^{T_\chi}$. Formally, for a suitable test function $f(\phi)$,
%This is a function which, when integrated against $f(\phi) \mathrm{d} \phi$ , gives the expectation value for the density of eigenvalues averaged over the ensemble $T_\chi$ with respect to a test function $f$. 
\begin{equation}
 R_{f}^{T_\chi}   =   \displaystyle \int _{T_{\chi}}  \sum _{j=1} ^{N}  f(\theta_{j})   \,  \mathrm{dA}_{T_{\chi}}
\end{equation}
where $\mathrm{dA}_{T_{\chi}} $ is the measure for the excised ensemble (which we shall write out explicitly later).  Using the argument principle, we may rewrite this in terms of the logarithmic derivative of the characteristic polynomial:
\begin{align}
 R_{f}^{T_\chi} = \int _{T_{\chi}} \frac{1}{2 \pi \mathrm{i}} \oint _{C}  \mathrm{i} \mathrm{e}^{\mathrm{i} \phi }\frac{\Lambda ' (\mathrm{e}^{\mathrm{i}\phi})}{\Lambda (\mathrm{e}^{\mathrm{i} \phi})} f(\phi)\mathrm{d}\phi \,  \mathrm{dA}_{T_{\chi}}.
\end{align}
Here, $C$ is a contour surrounding the segment of the real axis between $-\pi$ and $\pi$, and we have exploited the fact that all the eigenvalues of an orthogonal matrix lie on the unit circle in the complex plane. To continue, we first switch the order of integration:
\begin{equation}
\ R_{f}^{T_\chi} = \frac{1}{2 \pi   } \oint _{C}  \int _{T_{\chi}}  \frac{  \mathrm{e}^{\mathrm{i} \phi }\Lambda ' (\mathrm{e}^{\mathrm{i} \phi}  )}{\Lambda (\mathrm{e}^{\mathrm{i} \phi})} \,  f(\phi) \mathrm{dA}_{T_{\chi}} \,  \mathrm{d}\phi .
\end{equation}
Then, we note that the measure $\mathrm{d}A_{T_{\chi}}$ on $ T_{\chi}$ is equal to d$A$ (Haar measure on $SO(2N)$) multiplied by a Heaviside function:
\begin{equation}
 \ R_{f}^{T_\chi}= \frac{1}{2 \pi   }  \oint _{C}  \int _{SO(2N) }  H(\log \Lambda(1)- \chi) \frac{ \mathrm{e}^{\mathrm{i} \phi }\Lambda ' ( \mathrm{e}^{\mathrm{i} \phi})}{\Lambda ( \mathrm{e}^{\mathrm{i} \phi })} f(\phi) \mathrm{dA}  \, \mathrm{d}\phi.
\end{equation}
Let us explicitly choose the contour $C$ to be a rectangle of height $2 \epsilon$: 
\begin{align}
\ R_{f}^{T_\chi}= \ & \frac{1}{2 \pi   }  \left[  \int _{\pi}^{-\pi}  \int _{SO(2N)} H(\log  \Lambda(1)  - \chi)  \frac{  \mathrm{e}^{\mathrm{i} (\phi + \mathrm{i} \epsilon) }\Lambda ' (\mathrm{e}^{\mathrm{i} (\phi + \mathrm{i} \epsilon)})}{\Lambda (\mathrm{e}^{\mathrm{i} (\phi + \mathrm{i} \epsilon)})} f( \phi + \mathrm{i} \epsilon)  \,  \mathrm{dA}  \, \mathrm{d}\phi \   \right. \nonumber \\ 
& +   \int _{-\pi}^{\pi} \int _{SO(2N)} H(\log  \Lambda(1)  - \chi)  \frac{ \mathrm{e}^{\mathrm{i} (\phi - \mathrm{i} \epsilon) }\Lambda ' (\mathrm{e}^{\mathrm{i} (\phi - \mathrm{i} \epsilon) })}{\Lambda (\mathrm{e}^{\mathrm{i} (\phi - \mathrm{i} \epsilon) })} f(\phi - \mathrm{i} \epsilon )  \, \mathrm{dA}  \, \mathrm{d}\phi  \nonumber \\
& +   \int _{\epsilon }^{-\epsilon}  \int _{SO(2N)} H(\log  \Lambda(1)  - \chi)  \frac{  \mathrm{e}^{\mathrm{i} (-\pi+ \mathrm{i} \phi ) }\Lambda ' (\mathrm{e}^{\mathrm{i} (-\pi + \mathrm{i}\phi) })}{\Lambda (\mathrm{e}^{\mathrm{i} (-\pi+ \mathrm{i}\phi ) })} f(-\pi + \mathrm{i}\phi )  \, \mathrm{dA}  \, \mathrm{d}\phi  \nonumber \\
& + \left. \int _{-\epsilon }^{\epsilon}  \int _{SO(2N)} H(\log  \Lambda(1)  - \chi)  \frac{ \mathrm{e}^{\mathrm{i} (\pi+ \mathrm{i} \phi ) }\Lambda ' (\mathrm{e}^{\mathrm{i} (\pi + \mathrm{i}\phi) })}{\Lambda (\mathrm{e}^{\mathrm{i} (\pi+ \mathrm{i}\phi ) })} f(\pi + \mathrm{i}\phi )  \, \mathrm{dA}  \, \mathrm{d}\phi  \right] .
\end{align}
%As $\epsilon$ becomes small, the last two terms (from the vertical contours) will be $\mathcal{O}(\epsilon)$, so will be negligible compared with the first two terms which do not depend on $\epsilon$. We shall drop the $\mathcal{O}(\epsilon)$ terms,
If we restrict ourselves to test functions that are $2 \pi$-periodic, even, and analytic in a strip about the real axis, then the last two terms will cancel. Let us now rewrite the first two integrals so that they have all their poles on the same side of the contour (as this will allow us to move the contour onto the real line). To do so, we change variables on the first integral from $\phi \to -\phi $, then use the functional equation (\ref{eq:func}) (which takes $\phi - \mathrm{i} \epsilon \to  - \phi + \mathrm{i} \epsilon$) on the second to find that:
\begin{align}
&\ R_{f}^{T_\chi}= \nonumber \\
 &  \lim_{\epsilon \to 0^+}  \frac{1}{2 \pi  }   \int _{-\pi} ^{\pi} \int _{SO(2N)} H(\log  \Lambda(1)  - \chi) \, \left(-   \mathrm{e}^{ -\mathrm{i} (\phi - \mathrm{i} \epsilon) }  \frac{\Lambda ' (\mathrm{e}^{- \mathrm{i}( \phi- \mathrm{i} \epsilon)})}{\Lambda (\mathrm{e}^{ - \mathrm{i} (\phi - \mathrm{i} \epsilon) })} \right) f( \phi - \mathrm{i} \epsilon) \,  \mathrm{dA} \, \mathrm{d}\phi \nonumber  \\
& +   \frac{1}{2 \pi  }   \int _{-\pi} ^{\pi} \int _{SO(2N)} H(\log  \Lambda(1)  - \chi) \, \left(  2N -  \mathrm{e}^{ - \mathrm{i}(\phi - \mathrm{i} \epsilon)}  \frac{\Lambda ' (\mathrm{e}^{- \mathrm{i} (\phi - \mathrm{i} \epsilon)})}{\Lambda (\mathrm{e}^{- \mathrm{i}(\phi - \mathrm{i} \epsilon) })}  \right) f( \phi - \mathrm{i} \epsilon) \,  \mathrm{dA} \, \mathrm{d}\phi 
\end{align}
(making use of the fact that we have chosen $f$ to be even). The time has come to rewrite the Heaviside function in integral form using equation (\ref{eq:H}):
\begin{align}
\ R_{f}^{T_\chi}= &\   \lim_{\epsilon \to 0^+} \frac{1}{2 \pi  } \int _{-\pi} ^{\pi}   \int _{SO(2N)}\frac{  1 }{2 \pi \mathrm{i}} \int _{d-i \infty}^{d+i \infty} \frac{\exp(r \log  \Lambda(1)  - r \chi)}{r} f( \phi - \mathrm{i} \epsilon)   \, \nonumber \\
& \   \times   \left(   2N -   \mathrm{e}^{ - \mathrm{i}(\phi - \mathrm{i} \epsilon) }  \frac{\Lambda ' (\mathrm{e}^{-\mathrm{i}( \phi - \mathrm{i} \epsilon) })}{\Lambda (\mathrm{e}^{- \mathrm{i} ( \phi - \mathrm{i} \epsilon)})}  -   \mathrm{e}^{ - \mathrm{i}(\phi - \mathrm{i} \epsilon) }    \frac{\Lambda ' (\mathrm{e}^{ - \mathrm{i}( \phi - \mathrm{i} \epsilon )})}{\Lambda (\mathrm{e}^{ - \mathrm{i}( \phi - \mathrm{i} \epsilon)})}  \right)   \,  \mathrm{d}r \mathrm{dA} \, \mathrm{d}\phi \nonumber \\
=& \   \lim_{\epsilon \to 0^+}  \frac{1}{4 \pi^2 \text i }  \int _{-\pi} ^{\pi}   \int _{SO(2N)} \int _{d-i \infty}^{d+i \infty} \frac{ \mathrm{e}^{-\chi r} \Lambda(1)^r}{r}  f(\phi - \mathrm{i} \epsilon) \nonumber \\
&\times   \left(  2N    - 2    \mathrm{e}^{ -  \mathrm{i}(\phi- \mathrm{i} \epsilon ) }  \frac{\Lambda ' (\mathrm{e}^{ - \mathrm{i} (\phi -\mathrm{i} \epsilon) })}{\Lambda (\mathrm{e}^{ - \mathrm{i} (\phi - \mathrm{i} \epsilon)  })}  \right) \, \mathrm{d}r  \mathrm{dA} \, \mathrm{d}\phi  \nonumber \\
= &\   \lim_{\epsilon \to 0^+}  \frac{1}{4 \pi^2  \text i  } \int  _{-\pi} ^{\pi} \int _{d-i \infty}^{d+i \infty} \frac{ \mathrm{e}^{-\chi r }}{r}   \,   f(\phi - \mathrm{i} \epsilon) \nonumber \\
& \times   \left( 2N   \int _{SO(2N)} \Lambda(1)^r \mathrm{dA}  - 2   \int _{SO(2N)}   \mathrm{e}^{ - \mathrm{i}(\phi - \mathrm{i} \epsilon )}  \frac{\Lambda(1)^r\Lambda ' (\mathrm{e}^{ - \mathrm{i} (\phi - \mathrm{i} \epsilon )})}{\Lambda (\mathrm{e}^{ - \mathrm{i} (\phi - \mathrm{i} \epsilon) })} \mathrm{dA}  \right)  \,  \, \mathrm{d}r  \mathrm{d}\phi .
\end{align}
We can do the integrals over $SO(2N)$ to next-to-leading-order in $N$ using (\ref{moment1}) and (\ref{moment2}), assuming they hold for complex $r$:
\begin{align}
\ R_{f}^{T_\chi}&=  \frac{1}{4 \pi^2 \text i   }  \lim_{\epsilon \to 0} \int _{-\pi}^{\pi}  \int _{d-i \infty}^{d+i \infty}   \frac{  \mathrm{e}^{-\chi r }}{r}  f(\phi - \mathrm{i} \epsilon)  \mathcal{V}(N,r)     \nonumber \\  
 &  \times \left( 2N    \left(1 + \frac{r(r-1)^2}{2N} +\mathcal{O}(N^{-2})    \right)   +  2  \mathcal{U}(N,r, \mathrm{i}(\phi  - \mathrm{i} \epsilon) )  \left(1 + \mathcal{O}(N^{-2})  \right)\right)   \,  \, \mathrm{d}r \mathrm{d}\phi.
\end{align}
Here we are assuming that any $r$-dependence in the error term will not cause the error term to become unmanageably large when we integrate it. This seems a reasonable assumption since this method ultimately gives us good agreement with numerical results (see Figure \ref{one_level_density_ntlo}). Note that $\frac{  \mathrm{e}^{-\chi r }}{r}    \mathcal{V}(N,r)  \mathcal{U}(N,r, \text i \phi )$ has no poles in $\phi$ on the real line (the poles from the $z$ functions at 0 cancel out), so we can take the $\epsilon \to 0$ limit:
\begin{align}
\ R_{f}^{T_\chi}=& \frac{1}{4 \pi^2   \text i   }  \int _{-\pi}^{\pi}  \int _{d-i \infty}^{d+i \infty}   \frac{ \mathrm{e}^{-\chi r }}{r}   \mathcal{V}(N,r)  \nonumber \\
& \times \left(2N    \left(1 + \frac{r(r-1)^2}{2N}     \right)      + 2   \mathcal{U}(N, r,  \mathrm{i} \phi  )   \right) \left(1 + \mathcal{O}\left(N^{-2} \right)\right) \,  \, \mathrm{d}r   f(\phi )  \mathrm{d}\phi .
\end{align}

We can write this as a series of residues in $r$:
\begin{equation}\label{residues}
R_{f}^{T_\chi} =  \frac{2 \pi \text i}{4 \pi^2 \text i }  \int _{-\pi}^{\pi}  \sum _{\text{residues} \  r} R_{r} (\phi) f(\phi) \mathrm{d} \phi .
\end{equation}
% \frac{1}{2 \pi}    \mathbb{R}\mathrm{e} \left[ \int_{-\pi}^{\pi}  \left(  \displaystyle \sum_{k = 0}^{\infty} b_k \exp \left( (k+1/2)\chi \right) \right) f(\phi) \mathrm{d} \phi \right]
%where the $b_k$s are coefficients arising from the residues in $r$ which can be calculated using a computational algebra package such as Mathematica, 
We shall show that $R_0 (\phi)$ is the one level density for $SO(2N)$. This $R_0 (\phi)$ term comes from the real part of the leading order term in $N$ from the residue at $r=0$:
\begin{align}
 & \frac{1 }{2 \pi} \int _{-\pi}^{\pi}   \left[ \text{Residue} \left|    \frac{ \mathrm{e}^{-\chi r }}{r}   \mathcal{V}(N,r) \nonumber \right. \right. \\
 &\left. \left. \quad \quad \quad \quad \times  \left(2N \left(1 + \frac{r(r-1)^2}{2N}     \right)      + 2 \mathcal{U}(N, r,\mathrm{i} \phi )  \right) \,  \, \right| _{r=0}  \right]  f(\phi)  \mathrm{d}\phi    \nonumber \\
 = & \frac{1}{2\pi} \int _{-\pi}^{\pi}  \left[     \left( 2N  + 2 \mathcal{U}(N, 0, \mathrm{i}\phi) \right)    \right]f(\phi)\mathrm{d}\phi   \nonumber \\
= \ &\frac{1}{2\pi}\int _{-\pi}^{\pi}  \left[     2N - 2 z(-2 \mathrm{i} \phi) - 2  \mathrm{e}^{-2N \mathrm{i} \phi}z(2 \mathrm{i} \phi)     \right] f(\phi) \mathrm{d}\phi   \nonumber \\
= \ &\frac{1}{2\pi} \int _{-\pi}^{\pi} \left[   2N - \frac{2}{1 - \mathrm{e}^{  2 \mathrm{i} \phi}}  - 2 \frac{\mathrm{e}^{-2N \mathrm{i} \phi }}{1 - \mathrm{e}^{ -2  \mathrm{i} \phi}}  \right]f(\phi)  \mathrm{d}\phi  \nonumber \\
%= \ &\frac{1}{2\pi}\int _{-\pi}^{\pi}  \left[   2N - \frac{2 \mathrm{e} ^{-\mathrm{i} \phi}}{\mathrm{e}^{ -\mathrm{i} \phi} - \mathrm{e}^{ \mathrm{i} \phi}}  - \frac{2   \mathrm{e}^{-(2N-1) \mathrm{i} \phi }}{\mathrm{e}^{  \mathrm{i} \phi} - \mathrm{e}^{ - \mathrm{i} \phi}}  \right]f(\phi)  \mathrm{d}\phi + \mathcal{O}\left(N^{-2} \right)\nonumber \\
= \ & \frac{1}{2\pi} \int _{-\pi}^{\pi}  \left[   2N - \frac {2(\cos(\phi) - \mathrm{i} \sin(\phi))}{- 2 \mathrm{i} \sin(\phi)} +  \right. \nonumber \\
 \ \quad \quad& \left. \frac{2 (\cos((2N-1)\phi)  + \mathrm{i} \sin((2N-1) \phi)  )}{2 \mathrm{i} \sin(\phi) }         \right] f(\phi) \text d \phi \nonumber \\
= \ & \frac{1}{2 \pi}\int _{-\pi}^{\pi} \left( 2N-1 + \sin \left( (2N-1)\phi \right)/\sin(\phi) \right) f(\phi) \mathrm{d}\phi   .
\end{align}
Note that this is equal to the one level density from the full $SO(2N)$ ensemble.
 
To see why the other residues take the form they do in (\ref{residues}), note that the higher residues will come from the poles of $\frac{G(r+1)\sqrt{\Gamma(2r+1)} }{\sqrt{G(2r +1) \Gamma(r+1)}}$. We know that $G(r+1)$ and $\Gamma(r+1)^{-1}$ are entire functions, so the poles will arise from $\Gamma(2r+1)^{1/2}G(2r+1)^{-1/2}$. For the following, note that a Puiseaux series in $r$ is just a Laurent series in $r^{1/m}$, where $m$ is some positive integer.

Firstly, note that away from the negative integers and half-integers, $G(2r+1)^{-1/2}$ cannot diverge, since $G(2r+1)$ will be non-zero, and $\Gamma(2r+1)$ is meromorphic with poles only at the negative half integers. This rules out the possibility of poles in $\frac{G(r+1)\sqrt{\Gamma(2r+1)} }{\sqrt{G(2r +1) \Gamma(r+1)}}$ everywhere in the complex plane except at points in the set $\{-n/2 \| n \in \mathbb{N} \} $. Let us consider what happens at these points.

The $G$-function $G(2r+1)$ has zeros at $r \in \{-n/2 \|n \in \mathbb{N} \}$, each one of multiplicity $n$. If we expand $G(2r+1)^{-1}$ as a Laurent series about $-n/2$, then, we get terms proportional to $(r+n/2)^{-n}, (r+n/2)^{-n+1},(r+n/2)^{-n+2},  \dots $ etc. So, when we expand $G(2r+1)^{-1/2}$ around $-n/2$, we get a Puiseaux series with terms proportional to $(r+n/2)^{-n/2}, (r+n/2)^{-n/2+ 1},(r+n/2)^{-n/2+ 2} \dots$ etc. Note that we do not get powers of $(r+ n/2)^{-n/2 + 1/2}, \ (r + n/2)^{-n/2 + 3/2}$ etc., since this series must square to the Laurent series for  $G(2r+1)^{-1}$. 

We also know that $\Gamma(2r+1)$ has poles at $\{-n/2 \|n \in \mathbb{N} \}  $. In the Laurent expansion of $\Gamma(2r + 1)$ around such a point $-n/2$, there are terms proportional to $(r+n/2)^{-1},(r+n/2)^0,(r+n/2)^{1}$ etc. This means that in the Puiseaux expansion of $\Gamma(2r + 1)^{1/2}$ around the pole at $-n/2$, we have terms proportional to $(r+n/2)^{-1/2},(r+n/2)^{1/2}, (r+n/2)^{3/2}$ etc. Again, we do not get integer powers in this series, since it must square to the Laurent series for $\Gamma(2r + 1)$.

Consider $\Gamma(2r+1)^{1/2} G(2r +1)^{-1/2 }$ at a negative integer $\{-m \| m \in \mathbb{N} \} $. The series expansion of this product at $-m$ will have terms proportional to $(r + m)^{-m - 1/2}, (r + m)^{- m + 1/2}, \dots$ etc, from multiplying the Puiseaux series together. This series will not have a $(r + m)^{-1}$ term, so will not contribute a residue to (\ref{residues}).

However, at the negative half-integers $\{-\frac{2n+1}{2} \|n \in \mathbb{N} \}$, when we consider the product $\Gamma(2r+1)^{1/2}G(2r+1)^{-1/2}$, the product of the two Puiseaux series will be a Laurent series with terms proportional to  $(r+\frac{2n+1}2)^{-n - 1},(r+\frac{2n+1}2)^{-n }, (r+\frac{2n+1}2)^{-n + 1}$ etc. This will have a term proportional to $(r+ n/2)^{-1}$, so we will get a residue contributing to (\ref{residues}). This completes our demonstration that the sum of residues in (\ref{residues}) can be written as $\displaystyle \sum_{k = 0}^{\infty} b_k \exp \left( (k+1/2)\chi \right)$. So, 
\begin{equation}
R_{f}^{T_\chi} = R_0 + \frac{1}{2 \pi}    \mathbb{R}\mathrm{e} \left[ \int_{-\pi}^{\pi}  \left(  \displaystyle \sum_{k = 0}^{\infty} b_k \exp \left( (k+1/2)\chi \right) \right) f(\phi) \mathrm{d} \phi \right] ,
\end{equation}
which is consistent with Theorem 1.3 from \cite{kn:dhkms12}.

Figure \ref{one_level_density_ntlo} shows that for suitably chosen $N$ and $\chi$, just using the first four residues gives an excellent approximation to the one level density. 

%%%%%%%%%%%%%%%%%%%%%%%%%%%%%%%%%%%%%%%%%%%%%%%%%%%%%
\section{A mixed moment  of elliptic curve $L$-functions} \label{sect:Lratios}

The equivalent quantity to the random matrix moment we calculated in the last section is
\begin{align}
\mathcal{R}_{E,X} (\phi,r ) := \displaystyle \sum _{\substack{0 < d \leq X \\ \omega_E \chi_d( -M) = 1}} \left[  \frac{ L_E ' (1/2 + \mathrm{i} \phi, \chi_d ) }{L_E(1/2 + \mathrm{i} \phi, \chi_d)  }  L_E ( 1/2,\chi_d)^r   \right].
\end{align}
We will obtain an expression for this by differentiating once the ratios conjecture below, choosing $K$ to be $r+1$. 

In \cite{kn:cfz2}, Conrey, Farmer and Zirnbauer derived the following ratios conjecture.  It is not stated explicitly in that paper, but is mentioned after their equation (6.31) as following from their Conjecture 5.3 and their Lemma 6.8 in analogy with (6.31). There are possibly some typos in that paper, so another resource is \cite{kn:massna16} where the conjecture is stated at Conjecture 5.1 using a slightly different set notation and giving some potential conditions on $\alpha$ and $\gamma$ including ${\rm Re}{\gamma}\gg\frac{1}{\log X}$. 

\begin{align} \label{comb_sum}
   \displaystyle \sum _{\substack{0 < d \leq X \\ \omega_E \chi_d( -M) = 1}} \left[  \frac{ \displaystyle \prod_{k = 1} ^K L_E (1/2 + \alpha_k, \chi_d)   }{L_E(1/2 +  \gamma, \chi_d)  }   \right] =  \mathcal{Q}_{E,X}( \alpha,\gamma) +  \mathcal{O}\left(X^{1/2 + \epsilon}\right),
\end{align}
where
\begin{align} \label{nested_integrals}
\mathcal{Q}_{E,X}(\alpha,\gamma) =   \frac{(-1)^{K(K-1)/2} 2^K }{K! (2\pi \mathrm{i})^K}  \displaystyle \sum_{\substack{ 0<d\leq X \\ \omega_E \chi_d( -M) = 1}}  \left(  \frac{M |d|^2 }{4 \pi^2 }\right)^{-\tfrac{1}{2}  \sum_{k = 1}^K  \alpha_k }  \nonumber \\
\times \oint   \left(  \frac{M |d|^2 }{4 \pi^2 }\right)^{ \tfrac{1}{2} \sum_{k = 1}^K \left( w_k  \right)}  \left(  \frac{H_{\alpha,\gamma}(w_1,\dots,w_K) \Delta(w_1^2,\dots,w_K^2)^2 \displaystyle \prod_{k=1}^K w_k}{\displaystyle \prod_{j=1} ^K\prod_{k = 1}^K   (w_j - \alpha_k)(w_j + \alpha_k)} \right)  \mathrm{d}w_1 \dots \mathrm{d}w_K,
\end{align}
with
\begin{align} \label{eq:H_def}
&H_{\alpha,\gamma}(w_1,\dots,w_K) \nonumber \\&=   Y_E(w_1,\dots, w_K; \gamma ) A_E(w_1,\dots, w_K; \gamma)   \prod_{k=1}^K  g\left( \frac{\alpha_k  -w_k} 2   \right) ,
\end{align}
\begin{align}
Y_E( \alpha; \gamma ) := \ & \frac{ \displaystyle \prod_{j < k \leq K} \zeta(1 + \alpha_j + \alpha_k) \zeta( 1 + 2 \gamma)}{\displaystyle \prod_{k = 1}^K \zeta( 1 + \alpha_k + \gamma)  }.
\end{align}
and
\begin{equation}
g(s):=\frac{\Gamma(1-s)}{\Gamma(1+s)}.
\end{equation}
Here, $\alpha = (\alpha_1,\dots,\alpha_K)$ and $\gamma$ is a scalar and the contours of integration enclose all the $\alpha$'s but avoid the singularities of $g(s)$.

The arithmetic factor $A_E(\alpha,\gamma) $ is a product over primes which is convergent where we need to use it.   It was first derived in \cite{kn:cfz2}   and is examined in some detail.  It is stated in general at equation (5.37) in that paper (they call it $A_{E(\mathcal{D})}(\alpha, \gamma)$) and examined further in Section 6.3, however there are possible typos to beware of.  We use the version in Mason and Snaith \cite{kn:massna16} in Conjecture 5.1 and translate here their set notation for  an example 
for the specific family of quadratic twists of the $L$-function associated with the elliptic curve $E_{11}$.  
\begin{eqnarray}
A_E(\alpha,\gamma)& : =&  \frac{ \displaystyle \prod_{1\leq j < k \leq K} (1 - 11^{-1 - \alpha_j - \alpha_k})   (1 - 11^{-1-2\gamma})  }{ \displaystyle \prod_{k = 1} ^K ( 1 - 11^{-1 - \alpha_k - \gamma})  }  \\
&& \times \left( \frac{(1 -\lambda(11) 11^{-1/2 - \gamma} )} { \displaystyle \prod _{k = 1}^K (1 - \lambda(11)11^{-1/2 - \alpha_k})   }   \right) \nonumber \\
&&\times  \displaystyle \prod_{p \neq 11}   \left[  \frac{ \displaystyle  \prod_{1\leq j < k \leq K} (1 - p^{-1 - \alpha_j - \alpha_k})   (1 - p^{-1-2\gamma}) }{ \displaystyle \prod_{k = 1} ^K ( 1 - p^{-1 - \alpha_k - \gamma})   } \right.
\nonumber \\
&& \left. \times \frac 1 {1 + \frac 1 p } \left( \frac 1 2 \frac{(1 - \lambda(p) p^{-1/2 - \gamma } +   p^{-1 - 2 \gamma} )} { \displaystyle\prod _{k = 1}^K (1 - \lambda(p)p^{- 1/2 - \alpha_k} +  p^{-1 - 2\alpha_k})   }  \right.   \right. \nonumber \\&& \left. 
\left. +  \frac 1 2 \frac{(1 + \lambda(p) p^{-1/2 - \gamma } +   p^{-1 - 2 \gamma} )} {\displaystyle \prod _{k = 1}^K (1 + \lambda(p) p^{- 1/2 - \alpha_k} +  p^{-1 - 2\alpha_k})    }  + \frac 1 p  \right) \right] .
\end{eqnarray}

If we start using $(\alpha_1,\ldots,\alpha_k) = (\alpha, 0,\dots,0)$:
\begin{align}
\mathcal{Q}_{E,X}(\alpha,\gamma) = \displaystyle \sum_{\substack{ 0<d\leq X \\ \omega_E \chi_d( -M) = 1}}   \exp \left(- \mathcal{N}_{d}    \alpha    \right) \frac{(-1)^{K(K-1)/2} 2^K }{K! (2\pi \mathrm{i})^K} \nonumber \\
\times  \oint   \exp \left( \mathcal{N}_{d}   \sum_{k = 1}^K  w_k    \right)  \left(  \frac{H_{\alpha,\gamma}(w_1,\dots,w_K) \Delta(w_1^2,\dots,w_K^2)^2 \displaystyle   \prod_{n = 1}^K w_n^{3 - 2K}}{   \prod_{n = 1}^K(w_n - \alpha)(w_n + \alpha)} \right)  \mathrm{d}w_1 \dots \mathrm{d}w_K. \nonumber 
\end{align}
This has exactly the same polar structure as (\ref{eq:presum}) - it behaves like the equivalent integral for $SO(2N)$. In the above we have  defined $\mathcal{N}_d=\log\left(\frac{\sqrt{M}d}{2\pi}\right)$ to make the comparison easier. The $\mathcal{N}$ associated with the largest $d$ in the sum, $\mathcal{N}_X$, can be thought to be equivalent to the matrix size $N$ in the random matrix version.  Note that in the random matrix case we found both the leading order term and the next to leading order term in $N$, while here, for simplicity, we shall just work with the leading order term in $\mathcal{N}_X$.  Acting in analogy to the random matrix case at (\ref{eq:presum}) we know that we can shrink the contours of integration onto the poles, with narrow necks connecting them. As we showed in Section \ref{sect:zero_contributions_even}, the integral will cancel on either side of the necks, leaving a sum over three circular contours for each $w$ variable. The only terms which contribute to this sum will have the residue from one $w_i$ evaluated at $\pm \alpha$, and the rest evaluated at 0. Using this fact, and in analogy with (\ref{scaled_sum_even}), we get
\begin{align}\label{eq:sumdd}
\mathcal{Q}_{E,X}(\alpha,\gamma) &= \displaystyle \sum_{\substack{ 0<d\leq X \\ \omega_E \chi_d( -M) = 1}}   \exp \left(- \mathcal{N}_{d}    \alpha    \right) \frac{(-1)^{K(K-1)/2} 2^K }{K! (2\pi \mathrm{i})^K}(K\mathcal{T}^E_++K\mathcal{T}^E_-),
\end{align}
where
\begin{eqnarray}
\mathcal{T}^E_\pm&&=\zeta(1+2\gamma)\oint_0\cdots \oint_0  \exp \left( \mathcal{N} _d \sum_{k = 1}^{K-1}  w_k    \right)  \nonumber \\
&&\times\left(  \frac{\prod_{1\leq j<k\leq{K-1}}\zeta(1+w_j+w_k) \Delta(w_1^2,\dots,w_{K-1}^2)^2 \displaystyle   \prod_{n = 1}^{K-1} w_n^{3 - 2K}}{  \prod_{k=1}^{K-1}\zeta(1+w_k+\gamma) \prod_{n = 1}^{K-1}(w_n - \alpha)(w_n + \alpha)} \right) \prod_{k=1}^{K-1}g\left(-\frac{w_k}{2}\right)\nonumber \\
&&\times\oint_{\pm\alpha}A_{E}(w_1,\ldots,w_K;\gamma)g\left(\frac{\alpha-w_K}{2}\right)\prod_{j=1}^{K-1}\left[\zeta(1+w_j+w_K)(w_K^2-w_j^2)^2\right]\nonumber \\
&&\qquad\times \frac{\exp(\mathcal{N}_{d}w_K)w_K^{3-2K}}{\zeta(1+w_K+\gamma)(w_K-\alpha)(w_K+\alpha)}dw_K \; dw_1\cdots dw_{K-1}.
\end{eqnarray}
Evaluating the $w_K$ integral:
\begin{eqnarray}
\mathcal{T}^E_\pm&&=\zeta(1+2\gamma)(2\pi i)\frac{\exp(\pm \mathcal{N}_d\alpha) (\pm \alpha)^{3-2K}g\left(-(\pm 1-1)\frac{\alpha}{2}\right)}{\zeta(1\pm \alpha +\gamma)(\pm2\alpha)}\oint_0\cdots \oint_0  \exp \left( \mathcal{N}_{d}  \sum_{k = 1}^{K-1}  w_k    \right)  \nonumber \\
&&\times\left(  \frac{\prod_{1\leq j<k\leq{K-1}}\zeta(1+w_j+w_k) \Delta(w_1^2,\dots,w_{K-1}^2)^2 \displaystyle   \prod_{n = 1}^{K-1} w_n^{3 - 2K}}{  \prod_{k=1}^{K-1}\zeta(1+w_k+\gamma) \prod_{n = 1}^{K-1}(w_n - \alpha)(w_n + \alpha)} \right) \nonumber \\
&&\times A_{E}(w_1,\ldots,w_{K-1}, \pm \alpha;\gamma)\prod_{k=1}^{K-1}g\left(-\frac{w_k}{2}\right)\prod_{j=1}^{K-1}\left[\zeta(1+w_j\pm \alpha)(\alpha^2-w_j^2)^2\right]\nonumber \\
&&\qquad\times  dw_1\cdots dw_{K-1}.
\end{eqnarray}

We are interested in the large conductor limit, so let us scale the $w$-variables by $\mathcal{N}_X = \log\left(\frac {\sqrt{M} X}  {2 \pi}\right)$, and as in (\ref{eq:rem})

\begin{eqnarray}
\mathcal{T}^E_\pm&&=\zeta(1+2\gamma)(2\pi i)\frac{\exp(\pm \mathcal{N}_d\alpha) (\pm \alpha)^{3-2K}g\left(-(\pm 1-1)\frac{\alpha}{2}\right)}{\zeta(1\pm \alpha +\gamma)(\pm2\alpha)}\oint_0\cdots \oint_0  \exp \left( \frac{ \mathcal{N}_{d} }{\mathcal{N}_X}   \sum_{k = 1}^{K-1}  w_k    \right)  \nonumber \\
&&\times\left(  \frac{\prod_{1\leq j<k\leq{K-1}}\zeta(1+\tfrac{w_j}{\mathcal{N}_X}+\tfrac{w_k}{\mathcal{N}_X}) \Delta(w_1^2,\dots,w_{K-1}^2)^2 \displaystyle   \prod_{n = 1}^{K-1} w_n^{3 - 2K}}{  \prod_{k=1}^{K-1}\zeta(1+\tfrac{w_k}{\mathcal{N}_X}+\gamma)} \right) \nonumber \\
&&\times A_{E}(\tfrac{w_1}{\mathcal{N}_X},\ldots,\tfrac{w_{K-1}}{\mathcal{N}_X}, \pm \alpha;\gamma)\prod_{k=1}^{K-1}g\left(-\frac{w_k}{2\mathcal{N}_X}\right)\prod_{j=1}^{K-1}\left[\zeta(1+\tfrac{w_j}{\mathcal{N}_X}\pm \alpha)(\big(\tfrac{w_j}{\mathcal{N}_X}\big)^2-\alpha^2)\right]\nonumber \\
&&\qquad\times  dw_1\cdots dw_{K-1}.
\end{eqnarray}

Keeping just the leading order term, we have: 

\begin{align}\label{eq:zeta1}
\zeta\left(1+ \frac{x}{\mathcal{N}_X}\right) &= \frac{\mathcal{N}_X}{x}+ \mathcal{O}\left(1\right) \\
\label{eq:zeta2} \zeta\left(1+ \alpha  + \frac{x}{\mathcal{N}_X} \right)  &= \zeta\left( 1+\alpha \right)     + \mathcal{O}\left(\mathcal{N}_X^{-1}\right)   \\
\label{eq:zeta3} \frac 1 {\zeta\left( 1+\gamma + \frac x N    \right)} &= \frac 1 {\zeta(1+\gamma)}  + \mathcal{O}\left(\mathcal{N}_X^{-1} \right) \\
\label{eq:zeta4} \left(\frac{w_n}{\mathcal{N}_X}\right)^2 - \alpha^2&= -\alpha^2 + \mathcal{O}(\mathcal{N}_X^{-2})\\
\label{eq:zeta5} A_{E}(\tfrac{w_1}{\mathcal{N}_X},\ldots,\tfrac{w_{K-1}}{\mathcal{N}_X}, \pm \alpha;\gamma)&=A_E(0,\ldots, 0,\pm \alpha;\gamma) + \mathcal{O}\left(\mathcal{N}_X^{-1}\right)  \\
\label{eq:zeta6} g\left(-\frac{w_k}{2\mathcal{N}_X}\right)&=1+ \mathcal{O}\left(\mathcal{N}_X^{-1}\right) .
\end{align}

Using these,
\begin{eqnarray}
\mathcal{T}^E_\pm&&=(-1)^{K-1}\mathcal{N}_X^{(K-1)(K-2)/2}\zeta(1+2\gamma)(\pi i)\frac{\exp(\pm \mathcal{N}_d\alpha)g\left(-(\pm 1-1)\frac{\alpha}{2}\right)}{\zeta(1\pm \alpha +\gamma)}\nonumber \\
&&\times\oint_0\cdots \oint_0  \exp \left( \frac{ \mathcal{N}_{d} }{\mathcal{N}_X}   \sum_{k = 1}^{K-1}  w_k    \right)\prod_{1\leq j<k\leq K-1} \left( \frac{1}{w_j+w_k}\right)  \Delta(w_1^2,\dots,w_{K-1}^2)^2  \nonumber \\
&&\times \frac{\zeta(1\pm \alpha)^{K-1}}{\zeta(1+\gamma)^{K-1} }A_E(0,\ldots,0,\pm \alpha;\gamma)\prod_{n=1}^{K-1} w_n^{3-2K}dw_n (1+\mathcal{O}(\mathcal{N}_X^{-1})).
\end{eqnarray}

The integral here is very nearly $\mathcal{M}^{SO(2N)}$ with the exception of the exponential. In analogy with Section \ref{sect:M} we write
\begin{eqnarray}
\mathcal{T}^E_\pm&&=(-1)^{K-1}\mathcal{N}_X^{(K-1)(K-2)/2}\zeta(1+2\gamma)(\pi i)\frac{\exp(\pm \mathcal{N}_d\alpha)g\left(-(\pm 1-1)\frac{\alpha}{2}\right)}{\zeta(1\pm \alpha +\gamma)}\nonumber \\
&&\qquad  \frac{\zeta(1\pm \alpha)^{K-1}}{\zeta(1+\gamma)^{K-1} }A_E(0,\ldots,0,\pm \alpha;\gamma)\;\; \mathcal{M}^d \;\; (1+\mathcal{O}(\mathcal{N}_X^{-1})),
\end{eqnarray}
with
\begin{eqnarray}
 \mathcal{M}^d&=&\oint_0\cdots \oint_0\prod_{1\leq j<l \leq K-1} (w_l^2-w_j^2)(w_l-w_j)  \exp \left( \frac{ \mathcal{N}_{d} }{\mathcal{N}_X}   \sum_{k = 1}^{K-1}  w_k    \right) \prod_{n=1}^{K-1} w_n^{3-2K}dw_n.
\end{eqnarray}
Compare this with (\ref{eq:Mvanddeterimant}).  In exactly the same way as at (\ref{eq:Mvanddeterimant}) we write
\begin{eqnarray}
&& \mathcal{M}^d=\nonumber\\
&&(K-1)! \left|\begin{array}{cccc}\scriptscriptstyle{\oint w_1^{3-2K}\exp\left(\frac{\mathcal{N}_d}{\mathcal{N}_X}w_1\right) dw_1} & \scriptscriptstyle{\oint w_1^{5-2K}\exp\left(\frac{\mathcal{N}_d}{\mathcal{N}_X}w_1\right) dw_1}& \cdots & \scriptscriptstyle{\oint w_1^{-1}\exp\left(\frac{\mathcal{N}_d}{\mathcal{N}_X}w_1\right) dw_1}\\ \vdots & \vdots & \ddots & \vdots\\\scriptscriptstyle{\oint w_{K-1}^{1-K}\exp\left(\frac{\mathcal{N}_d}{\mathcal{N}_X}w_{K-1}\right) dw_{K-1}} & \scriptscriptstyle{\oint w_{K-1}^{3-K}\exp\left(\frac{\mathcal{N}_d}{\mathcal{N}_X}w_{K-1}\right) dw_{K-1}}& \cdots & \scriptscriptstyle{\oint w_{K-1}^{K-3}\exp\left(\frac{\mathcal{N}_d}{\mathcal{N}_X}w_{K-1}\right) dw_{K-1}}\end{array}\right|\nonumber\\
&&=(K-1)! \left|\begin{array}{cccc}\scriptscriptstyle{\left(\tfrac{\mathcal{N}_X}{\mathcal{N}_d}\right)^{4-2K}\oint w_1^{3-2K}\exp\left(w_1\right) dw_1} & \cdots & \scriptscriptstyle{\oint w_1^{-1}\exp\left(w_1\right) dw_1}\\ \vdots  & \ddots & \vdots\\\scriptscriptstyle{\left(\tfrac{\mathcal{N}_X}{\mathcal{N}_d}\right)^{2-K}\oint w_{K-1}^{1-K}\exp\left(w_{K-1}\right) dw_{K-1}} & \cdots & \scriptscriptstyle{\left(\tfrac{\mathcal{N}_X}{\mathcal{N}_d}\right)^{K-2}\oint w_{K-1}^{K-3}\exp\left(w_{K-1}\right) dw_{K-1}}\end{array}\right|\nonumber\\
&&=(2\pi i)^{K-1}(K-1)! \left(\frac{\mathcal{N}_d}{\mathcal{N}_X}\right)^{(K-1)(K-2)/2}\\
&&\qquad \times\det_{(K-1)\times(K-1)}\left(\begin{array}{ccccc} \frac{1}{\Gamma(2K-3)} &  \frac{1}{\Gamma(2K-5)} &  \frac{1}{\Gamma(2K-7)} &\cdots & \frac{1}{\Gamma(1)} \\ \frac{1}{\Gamma(2K-4)}& \frac{1}{\Gamma(2K-6)}& \frac{1}{\Gamma(2K-8)}& \cdots & \frac{1}{\Gamma(0)} \\ \vdots &\vdots &\vdots &\ddots&\vdots \\  \frac{1}{\Gamma(K-1)}&  \frac{1}{\Gamma(K-3)}&  \frac{1}{\Gamma(K-5)} &\cdots &  \frac{1}{\Gamma(3-K)}\end{array}\right)\nonumber\\
&&=\left(\frac{\mathcal{N}_d}{\mathcal{N}_X}\right)^{(K-1)(K-2)/2}\mathcal{M}^{SO(2N)},
 \end{eqnarray}
 by comparison with (\ref{eq:detforM}).

Returning to (\ref{eq:sumdd})
\begin{eqnarray}
\mathcal{Q}_{E,X}(\alpha,\gamma) &&= \displaystyle \sum_{\substack{ 0<d\leq X \\ \omega_E \chi_d( -M) = 1}}   \exp \left(- \mathcal{N}_{d}    \alpha    \right) \frac{(-1)^{K(K-1)/2} 2^K }{K! (2\pi \mathrm{i})^K}(K\mathcal{T}^E_++K\mathcal{T}^E_-)\\
&&= \displaystyle \sum_{\substack{ 0<d\leq X \\ \omega_E \chi_d( -M) = 1}}   \exp \left(- \mathcal{N}_{d}    \alpha    \right) \frac{(-1)^{K(K-1)/2} 2^K }{(K-1)! (2\pi \mathrm{i})^K}\nonumber \\
&& \times\left(\frac{\mathcal{N}_d}{\mathcal{N}_X}\right)^{(K-1)(K-2)/2} \mathcal{M}^{SO(2N)}(-1)^{K-1}\mathcal{N}_X^{(K-1)(K-2)/2}(\pi i)\nonumber \\
&& \times \Bigg( \frac{\exp( \mathcal{N}_d\alpha)\zeta(1+2\gamma)}{\zeta(1+\gamma)^{K-1}\zeta(1+ \alpha +\gamma)}  \zeta(1+ \alpha)^{K-1}A_E(0,\ldots,0, \alpha;\gamma) \nonumber \\
&& + \frac{\exp(-\mathcal{N}_d\alpha)\zeta(1+2\gamma)g\left(\alpha\right)}{\zeta(1+\gamma)^{K-1}\zeta(1- \alpha +\gamma)}  \zeta(1- \alpha)^{K-1}A_E(0,\ldots,0,- \alpha;\gamma)\Bigg) (1+\mathcal{O}(\mathcal{N}_X^{-1})).\nonumber
\end{eqnarray}
Using (\ref{eq:barnesGforM}) for $\mathcal{M}^{SO(2N)}$,
\begin{eqnarray}
\mathcal{Q}_{E,X}(\alpha,\gamma) &&= \displaystyle \sum_{\substack{ 0<d\leq X \\ \omega_E \chi_d( -M) = 1}} 2^{(K-1)^2/2}\mathcal{N}_d^{(K-1)(K-2)/2}\frac{G(K)\sqrt{\Gamma(2K-1)}}{\sqrt{G(2K-1)\Gamma(K)}}\nonumber \\
&& \times \Bigg( \frac{\zeta(1+2\gamma)}{\zeta(1+\gamma)^{K-1}\zeta(1+ \alpha +\gamma)}  \zeta(1+ \alpha)^{K-1}A_E(0,\ldots,0, \alpha;\gamma) \nonumber \\
&& + \frac{\exp(-2\mathcal{N}_d\alpha)\zeta(1+2\gamma)\Gamma(1-\alpha)}{\zeta(1+\gamma)^{K-1}\zeta(1- \alpha +\gamma)\Gamma(1+\alpha)}  \zeta(1- \alpha)^{K-1}A_E(0,\ldots,0,- \alpha;\gamma)\Bigg) (1+\mathcal{O}(\mathcal{N}_X^{-1})).\nonumber
\end{eqnarray}

We set $K=r+1$, write $A_E(0,\ldots,0,\alpha;\gamma)$ as $\tilde{A}_E(\alpha,\gamma)$ for simplicity,  and differentiate this with respect to $\alpha$:
\begin{eqnarray}
\frac{d}{d\alpha} \mathcal{Q}_{E,X}(\alpha,\gamma)&=&2^{r^2/2}  \displaystyle \sum_{\substack{ 0<d\leq X \\ \omega_E \chi_d( -M) = 1}}\mathcal{N}_d ^{\;r(r-1)/2} \frac{\zeta(1 + 2\gamma)}{\zeta(1 + \gamma)^r} \left(\frac{d}{d\alpha}\tilde{A}_E(\alpha,\gamma) \frac{ \zeta(1 + \alpha)^r }{\zeta(1 + \alpha + \gamma)} \right.   \nonumber \\ 
&& \left.  + \tilde{A}_E(\alpha,\gamma) \frac{ r \zeta(1 + \alpha)^{r-1} \zeta'(1 + \alpha) }{\zeta(1 + \alpha + \gamma)}        \right.  \nonumber \\
&&-  \tilde{A}_E(\alpha,\gamma) \frac{\zeta(1 + \alpha)^r \zeta'(1 + \alpha + \gamma) }{\zeta(1 + \alpha + \gamma)^2 }\nonumber \\
&&+ \frac{d}{d\alpha}\tilde{A}_E(- \alpha,\gamma) \mathrm{e}^{- 2\mathcal{N}_d \alpha}\frac{\Gamma(1- \alpha )}{\Gamma(1+\alpha )} \frac{\zeta( 1 - \alpha)^r}{\zeta( 1 - \alpha + \gamma) }   \nonumber \\
&&- \tilde{A}_E(-\alpha,\gamma) \mathrm{e}^{- 2\mathcal{N}_d \alpha}\frac{ \Gamma(1 - \alpha)}{\Gamma(1 + \alpha)} \left(\Psi(1 - \alpha) + \Psi(1 + \alpha) \right)  \frac{\zeta(1 - \alpha)^r}{\zeta( 1 - \alpha + \gamma) } \nonumber \\
&&- \tilde{A}_E(-\alpha,\gamma) \frac{\Gamma(1 - \alpha)}{\Gamma(1+ \alpha )} \mathrm{e}^{- 2\mathcal{N}_d \alpha} \frac{r \zeta( 1 - \alpha)^{r-1} \zeta'(1 - \alpha)      }{\zeta( 1- \alpha + \gamma)} \nonumber \\
&& -2\mathcal{N}_d  \tilde{A}_E(-\alpha,\gamma) \frac{\Gamma(1 - \alpha )}{\Gamma(1+ \alpha)} \mathrm{e}^{- 2\mathcal{N}_d \alpha} \frac{ \zeta( 1 - \alpha)^{r}     }{\zeta( 1- \alpha + \gamma)} 
 \nonumber \\
&&\left. + \tilde{A}_E(-\alpha,\gamma) \frac{\Gamma(1 - \alpha )}{\Gamma(1 + \alpha )} \mathrm{e}^{- 2\mathcal{N}_d \alpha} \frac{ \zeta( 1 - \alpha)^{r} \zeta'(1 - \alpha + \gamma )      }{\zeta( 1- \alpha + \gamma)^2}\right) \nonumber \\
&&\times \frac{G(r+1) \sqrt{\Gamma(2r+1)}}{\sqrt{G(2r+1) \Gamma(r+1)}}  \left(1  + \mathcal{O}\left(\mathcal{N}_X^{-1} \right)\right) . 
\end{eqnarray}
Here $\Psi(x)=\Gamma'(x)/\Gamma(x)$. Fortunately, when we set $\alpha = \gamma =  \mathrm{i} \phi$ (note $\phi$ is not necessarily real), several of these terms disappear.  We note that $\lim _{\alpha \to \gamma}\zeta'(1- \alpha + \gamma)/\zeta(1 - \alpha + \gamma)^2 = 1$. Also $\lim_{\alpha \to \gamma}1/\zeta(1-\alpha+\gamma)=0$. 
%\begin{eqnarray}
%A_E(\alpha,\gamma) &=&  \frac{ \displaystyle  (1 - 11^{-1 - \alpha})^{r}  (1 - 11^{-1 }) ^{r(r-1)/2}   }{ \displaystyle ( 1 - 11^{-1 - \alpha - \gamma})  ( 1 - 11^{-1  - \gamma})^{r}    (1 - %11^{-1-2\gamma})     } \nonumber \\
%&&\times \frac{1}{(1+1/11)^{r+1}}\left( \frac{1}{2}\frac{(1-11^{-1-\gamma})}{(1-11^{-1-\alpha})}+\frac{1}{2}\frac{(1+11^{-1-\gamma})}{(1+11^{-1-\alpha})}+\frac{1}{11}\right)  \nonumber \\
%&&\times \displaystyle \prod_{p \neq 11}  \frac{(1 - p^{-1 - \alpha})^{r} (1 - p^{-1})^{r(r-1)/2 }  }{  ( 1 - p^{-1 - \alpha- \gamma })( 1 - p^{-1 - \gamma})^{r}  (1 - p^{-1-2\gamma})    } %\nonumber \\
%&&\times \frac 1 {1 + \frac 1 p } \left( \frac 1 2 \frac{(1 - \lambda(p) p^{-1/2 - \gamma } +   p^{-1 - 2 \gamma} )} { (1 -  p^{- 1/2 - \alpha} + \lambda(p)p^{-1 - 2\alpha})  (1 - p^{- 1/2} +  \lambda(p)p^{-1})^{r}    }    \right. \nonumber \\
%&&\left. +  \frac 1 2 \frac{(1 + \lambda(p) p^{-1/2 - \gamma } +   p^{-1 - 2 \gamma} )} { (1 + \lambda(p) p^{- 1/2 - \alpha} +  \lambda(p)p^{-1 - 2\alpha}) (1 + \lambda(p) p^{- 1/2 } + \lambda(p) p^{-1})^{r}   }. + \frac{1}{p} \right) .
%\end{eqnarray}

%so $A_E(\mathrm{i} \phi,\mathrm{i} \phi) = 1$.
\begin{align}
& \displaystyle \sum _{\substack{0 < d \leq X \\ \omega_E \chi_d( -M) = 1}} \left[ \frac{ L_E ' (1/2 + \mathrm{i} \phi, \chi_d ) }{L_E(1/2 +  \mathrm{i} \phi, \chi_d)  }  L_E ( 1/2,\chi_d)^r   \right]   \\
&=  \displaystyle \sum _{\substack{0 < d \leq X \\ \omega_E \chi_d( -M) = 1}}  2^{r^2/2} \mathcal{N}_d ^{r(r-1)/2} \frac{G(r+1) \sqrt{\Gamma(2r + 1)}}{\sqrt{G(2r+1) \Gamma(r+1) }}  \left(\tilde{A}_E^1(\mathrm{i} \phi) + \tilde{A}_E(\mathrm{i} \phi,\mathrm{i} \phi) \frac{ r   \zeta'(1 + \mathrm{i} \phi) }{\zeta( 1+ \mathrm{i}  \phi) }        \right.  \nonumber \\
&\left. - \tilde{A}_E(\mathrm{i} \phi,\mathrm{i} \phi)  \frac{\zeta'(1 +2 \mathrm{i} \phi ) }{\zeta(1 + 2\mathrm{i} \phi )} + \frac{\zeta(1 + 2\mathrm{i} \phi) \zeta( 1 -\mathrm{i}  \phi)^r }{\zeta(1 +\mathrm{i}  \phi)^r}  \tilde{A}_E(-\mathrm{i} \phi,\mathrm{i} \phi) \frac{\Gamma(1 - \mathrm{i} \phi )}{\Gamma(1 + \mathrm{i} \phi )} \mathrm{e}^{- 2 \mathrm{i} \mathcal{N}_d \phi}      \right) \left(1  + \mathcal{O}\left(\mathcal{N}_X^{-1} \right)\right).\nonumber
\end{align}
Here, $\tilde{A}_E^1 (\phi)=  d /  d \alpha \tilde{A}_E(\alpha,\gamma)|_{\alpha = \gamma = \phi}$. Now let 
\begin{align}
 \mathcal{V}(\mathcal{N}_d,r) :=  2^{r^2/2} \mathcal{N}_d ^{r(r-1)/2} \frac{G(r+1) \sqrt{\Gamma(2r + 1)}}{\sqrt{G(2r+1) \Gamma(r+1) }} 
\end{align}
and 
\begin{eqnarray}
  \mathcal{U}_E(\mathcal{N}_d,r,\phi) &:=&\tilde{A}_E^1(\mathrm{i} \phi) + \tilde{A}_E(\mathrm{i} \phi,\mathrm{i} \phi) \frac{ r   \zeta'(1 + \mathrm{i} \phi) }{\zeta( 1+ \mathrm{i}  \phi) }       - \tilde{A}_E(\mathrm{i} \phi,\mathrm{i} \phi)  \frac{\zeta'(1 +2 \mathrm{i} \phi ) }{\zeta(1 + 2\mathrm{i} \phi )} \\
  &&\qquad+ \frac{\zeta(1 + 2\mathrm{i} \phi) \zeta( 1 -\mathrm{i}  \phi)^r }{\zeta(1 +\mathrm{i}  \phi)^r}  \tilde{A}_E(-\mathrm{i} \phi,\mathrm{i} \phi) \frac{\Gamma(1 - \mathrm{i} \phi )}{\Gamma(1 + \mathrm{i} \phi )} \mathrm{e}^{- 2 \mathrm{i} \mathcal{N}_d \phi} \nonumber
\end{eqnarray} 
so
\begin{align}\label{eq:finalLmoment}
\displaystyle \sum _{\substack{0 < d \leq X \\ \omega_E \chi_d( -M) = 1}} \left[ \frac{ L_E ' (1/2 + \mathrm{i} \phi, \chi_d ) }{L_E(1/2 +  \mathrm{i} \phi, \chi_d)  }  L_E ( 1/2,\chi_d)^r   \right]  \nonumber \\
= \displaystyle \sum _{\substack{0 < d \leq X \\ \omega_E \chi_d( -M) = 1}} \mathcal{V}(\mathcal{N}_d,r)  \mathcal{U}_E(\mathcal{N}_d,r,\phi)  \left(1  + \mathcal{O}\left(\mathcal{N}_X^{-1} \right)\right).
\end{align}

%%%%%%%%%%%%%%%%%%%%%%%%%%%%%%%%%%%%%%%%%%%%%%%%%
\section{One level density for quadratic twists of elliptic curve $L$-functions} \label{sect:Lonelevel}

In Section \ref{sect:old_even}, we used both the leading order term and the next to leading order terms (in $N$) for the mixed moment in order to calculate the one level density over $SO(2N)$. However, to make the calculation easier, we shall only use the leading order term in $\mathcal{N}_X$ this time. Let $\gamma_d$ denote a zero of $L_E(s ,\chi_d)$ on the half line. For a test function $f$, let us consider the one level density
\begin{align}
 \displaystyle \sum _{\substack{0 < d \leq X \\ \omega_E \chi_d(-M) = +1 }} \displaystyle \sum_{\gamma_d} f(\gamma_d),
\end{align}
where $f$ is a suitable test function (e.g. an even Schwartz function). 
Using the argument principle, we find
\begin{align}
\displaystyle \sum _{\substack{0 < d \leq X \\ \omega_E \chi_d(-M) = +1 }} \displaystyle \sum_{\gamma_d} f(\gamma_d)=  \displaystyle \sum _{\substack{0 < d \leq X \\ \omega_E \chi_d(-M) = +1 }} \frac 1 {2 \pi \mathrm{i}  } \left( \int _{(c)} - \int_{(1 - c)} \right)  \frac{L'_E(s,\chi_d)}{L_E(s,\chi_d)} f(-i(s-1/2)) \mathrm{d} s     ,       
\end{align}
where $(c)$ denotes a vertical line from $c - \mathrm{i} \infty$ to $c + \mathrm{i} \infty$, and $3/4 > c > 1/2 + 1/\log X$. One can continue from here by replacing the ratio of $L$-functions with the result of the ratios conjecture for  $\displaystyle \sum _{\substack{0 < d \leq X \\ \omega_E \chi_d(-M) = +1 }} \frac{L'_E(s,\chi_d)}{L_E(s,\chi_d)} $. This was done in \cite{kn:hks}, and results in an excellent prediction for the one level density away from the critical point $s=1/2$, but which does not capture the extra zero repulsion at the origin, as can be seen in Figure  \ref{fig:E11_zero_repulsion}. Based on the work done in \cite{kn:dhkms12} and in \cite{kn:coomorsna}, we believe that this extra repulsion is linked to the discretisation of the central values $L_E(1/2,\chi_d)$ due to the central value formula (\ref{eq:kz_bound}). That is, the central value is either zero or it is greater than $\kappa_E/|d|^{1/2}$. Therefore, we might expect that if we explicitly enforce this bound using a Heaviside step function, we would get a prediction for the one level density which does contain information about the central value which might have been lost in the approximations of the ratios conjecture recipe.  Therefore we define our one level density in the following way:
\begin{eqnarray}
&&S_1(f) =   \\
  &&\displaystyle \sum _{\substack{0 < d \leq X \\ \omega_E \chi_d(-M) = +1 }} \frac 1 {2 \pi \mathrm{i}  } \left(\int _{(c)} - \int_{(1 - c)}\right)  \frac{L'_E(s,\chi_d)}{L_E(s,\chi_d)} H\left( \log(L_E(1/2,\chi_d) - \xi )\right) f(-i(s-1/2)) \mathrm{d} s\nonumber
\end{eqnarray}
where $\xi =  \log(\kappa_E /|d|^{1/2})$. The value of $\kappa_E$ has been computed for many families by Michael Rubinstein and can be found in Table 3 of \cite{kn:ckrs05}. Using the integral representation of the Heaviside function (\ref{eq:H}), we get
\begin{eqnarray}
&&S_1(f) =  \\
  &&\displaystyle \sum _{\substack{0 < d \leq X \\ \omega_E \chi_d(-M) = +1 }} \frac {-1} {4 \pi^2} \left( \int _{(c)} - \int_{(1 - c)} \right)  \frac{L'_E(s,\chi_d)}{L_E(s,\chi_d)} \int _{(b)} L_E(1/2, \chi_d)^r  \frac{ \mathrm{e}^{-\xi r} }{r} \mathrm{d} r    f(-i(s-1/2)) \mathrm{d} r \mathrm{d} s\nonumber
\end{eqnarray}
where $b>0$. We can move the logarithmic derivative of $L$ inside the $r$ integral:
\begin{align}
S_1(f) =  \displaystyle \sum _{\substack{0 < d \leq X \\ \omega_E \chi_d(-M) = +1 }} \frac {-1} {4 \pi^2 } \left(  \int _{(c)} - \int_{(1 - c)} \right)   \left[  \int _{(b)} \frac{L'_E(s,\chi_d)}{L_E(s,\chi_d)}  L_E(1/2, \chi_d)^r  \frac{  \mathrm{e}^{-\xi r}}{r} \mathrm{d} r    \right]   \nonumber \\
\times   f(-i(s-1/2))   \mathrm{d} s.
\end{align}
Let us first consider the integral over $s$ on the $c$ line. If we replace the ratios of $L$-functions by the result in (\ref{eq:finalLmoment}), the integrand no longer has poles in $s$ for $\mathbb{I}\mathrm{m}[s] >0$ on the $1/2$ line, so we can then move the path of integration onto $\mathbb{R}\mathrm{e}[s] = 1/2$ to get
%\begin{align}  
%\displaystyle \sum _{\substack{0 < d \leq X \\ \omega_E \chi_d(-M) = +1 }} \frac 1 {4\pi^2} \int _{- \infty}^{\infty}  \left[  \int _{(b)} 2^{r^2} (\mathcal{N}_X) ^{r(r-1)/2} \frac{G(r+1) \sqrt{\Gamma(2r + 1)}}{\sqrt{G(2r+1) \Gamma(r+1) }}   \right. \nonumber \\
% \left(A_E'(\mathrm{i} \phi,\mathrm{i} \phi) +  \frac{ r   \zeta'(1 + \mathrm{i} \phi) }{\zeta( 1+  \mathrm{i} \phi) }       
%-   \frac{\zeta'(1 +2 \mathrm{i} \phi ) }{\zeta(1 + 2\mathrm{i} \phi )}   \right. \nonumber \\
% \left. - \frac{\zeta(1 + 2\mathrm{i} \phi) \zeta( 1 - \mathrm{i} \phi)^r }{\zeta(1 + \mathrm{i} \phi)^r} \left. A_E(-\mathrm{i} \phi,\mathrm{i} \phi) \frac{\Gamma(1 - \mathrm{i} \phi)}{\Gamma(1 + \mathrm{i} \phi)} \mathrm{e}^{- 2\mathcal{N}_X \mathrm{i} \phi}      \right)  \frac{\mathrm{e}^{-\xi r}}{r} \mathrm{d} r    f(\phi)  \mathrm{d} r   \right]    \mathrm{d} \phi
%\end{align}
%where $s = 1/2 + \mathrm{i} \phi$. Let 
%\begin{align}
%z_E(X,r,\phi) :=  \displaystyle \sum _{\substack{0 < d \leq X \\ \omega_E \chi_d(-M) = +1 }}  
%\left(A_E'(\phi,\phi) +  \frac{ r   \zeta'(1 + \phi) }{\zeta( 1+  \phi) }       
%-   \frac{\zeta'(1 +2 \phi ) }{\zeta(1 + 2\phi )}   \right. \nonumber \\
% \left. - \frac{\zeta(1 + 2\phi) \zeta( 1 - \phi)^r }{\zeta(1 + \phi)^r}  A_E(-\phi,\phi) \frac{\Gamma(1 - \phi )}{\Gamma(1 + \phi)} \mathrm{e}^{- 2\mathcal{N}_X \phi}      \right) ,
%\end{align}
%and 
%\begin{align}
%\mathcal{V}(N,r) := 2^{r^2} (\mathcal{N}_X) ^{r(r-1)/2} \frac{G(r+1) \sqrt{\Gamma(2r + 1)}}{\sqrt{G(2r+1) \Gamma(r+1) }}   
%\end{align}
%so the $(c)$ integral can be written as
\begin{align}
 \displaystyle \sum _{\substack{0 < d \leq X \\ \omega_E \chi_d(-M) = +1 }}\frac {1 } {4 \pi^2 \mathrm{i}} \int _{- \infty}^\infty  \left[  \int _{(b)} \mathcal{V}(\mathcal{N}_d,r) \mathcal{U}_E(\mathcal{N}_d,r,\mathrm{i}  \phi)   \left(1  + \mathcal{O}\left(\mathcal{N}_X^{-1} \right)\right) \frac{ \mathrm{e}^{-\xi r}} r \mathrm{d} r    f(\phi )  \mathrm{d} r   \right]    \mathrm{d} \phi .
\end{align}
Here again we are assuming $r$ can be a complex variable, rather than just an integer, and from here on we will neglect error terms. Our purpose is to show that this method gives qualitatively the correct behaviour of the one level density near the origin. 

For the $(1-c)$ integral, we change variables $s \to 1 -s$ so that we can write it as an integral over $(c)$:
\begin{align}
\displaystyle \sum _{\substack{0 < d \leq X \\ \omega_E \chi_d(-M) = +1 }} \frac {1} {4 \pi^2 }  \int_{(c)} \left[  \int _{(b)} \frac{L'_E(1 - s,\chi_d)}{L_E(1 - s,\chi_d)}  L_E(1/2, \chi_d)^r  \frac{  \mathrm{e}^{-\xi r}}{r} \mathrm{d} r    \right]    f(i(s-1/2))   \mathrm{d} s.
\end{align}
The functional equation for these $L$-functions looks like
\begin{equation}
L_E(1/2+i\phi,\chi_d)=\mathcal{X}(1/2+i\phi,\chi_d)L_E(1/2-i\phi,\chi_d),
\end{equation}
where
\begin{align}
\mathcal X(1/2 + \mathrm{i} \phi,\chi_d) := \exp( - 2 \mathcal{N}_d \mathrm{i} \phi )  \frac{\Gamma(1 - \mathrm{i} \phi)}{\Gamma(1 +\mathrm{i}  \phi)}.
\end{align}

We take the derivative of the functional equation
\begin{align}
\frac{L_E '}{L_E}(1/2 - \mathrm{i} \phi, \chi_d)  = \frac{\mathcal X'}{\mathcal X} (1/2 + \mathrm{i}   \phi, \chi_d) - \frac{L_E'}{L_E} (1/2 + \mathrm{i} \phi,\chi_d).
\end{align}
We then replace the ratio of $L$-functions with the result in (\ref{eq:finalLmoment}), to find that for the integral over $(1 - c)$, we get:
\begin{align}
\approx\displaystyle \sum _{\substack{0 < d \leq X \\ \omega_E \chi_d(-M) = +1 }} \frac {1} {4 \pi^2 \mathrm{i} } \int _{-\infty} ^\infty  \left[  \int _{(b)}  \left(   \frac{\mathcal{X}'}{\mathcal{X}}(1/2 + \mathrm{i} \phi,\chi_d) -   \mathcal{U}_E(\mathcal{N}_d,r,\mathrm{i} \phi)    \right) \right. \nonumber \\
\left. \times  \mathcal{V}(\mathcal{N}_d,r)    \frac{\mathrm{e}^{-\xi r}}{r}   f(\phi)  \mathrm{d} r   \right]    \mathrm{d} \phi .
\end{align}
We know from  Section \ref{sect:old_even} that $ \mathcal{V}(\mathcal{N}_d,r) $ has poles at the negative half-integers, and the $r=0$ term is the one level density already derived in \cite{kn:hks}, so we can write this as a series of residues:
\begin{align}
\label{eq:excised_density_prediction}
S_1(f) \approx \displaystyle \sum_{j = 0,-1/2,-3/2,\dots} \sum _{\substack{0 < d \leq X \\ \omega_E \chi_d(-M) = +1 }} \frac {1} {2 \pi } \int _{-\infty} ^\infty \text{Res}\left[ \left( -   \frac{\mathcal{X}'}{\mathcal{X}}(1/2 + \mathrm{i} \phi, \chi_d) +  2 \mathcal{U}_E(\mathcal{N}_d,r,\mathrm{i} \phi)   \right) \right.\nonumber \\
\left. \times \mathcal{V}_E(\mathcal{N}_d,r)    \frac{\mathrm{e}^{-\xi r}}{r}    f(\phi)  \right]_{r = j}  \mathrm{d} \phi .
\end{align}
%%%%%%%%%%%%%%%%%%%%%%%%%%%%%%%%%

\subsection{The $r=0$ term}
\begin{figure}
\centering
\includegraphics[width=10cm]{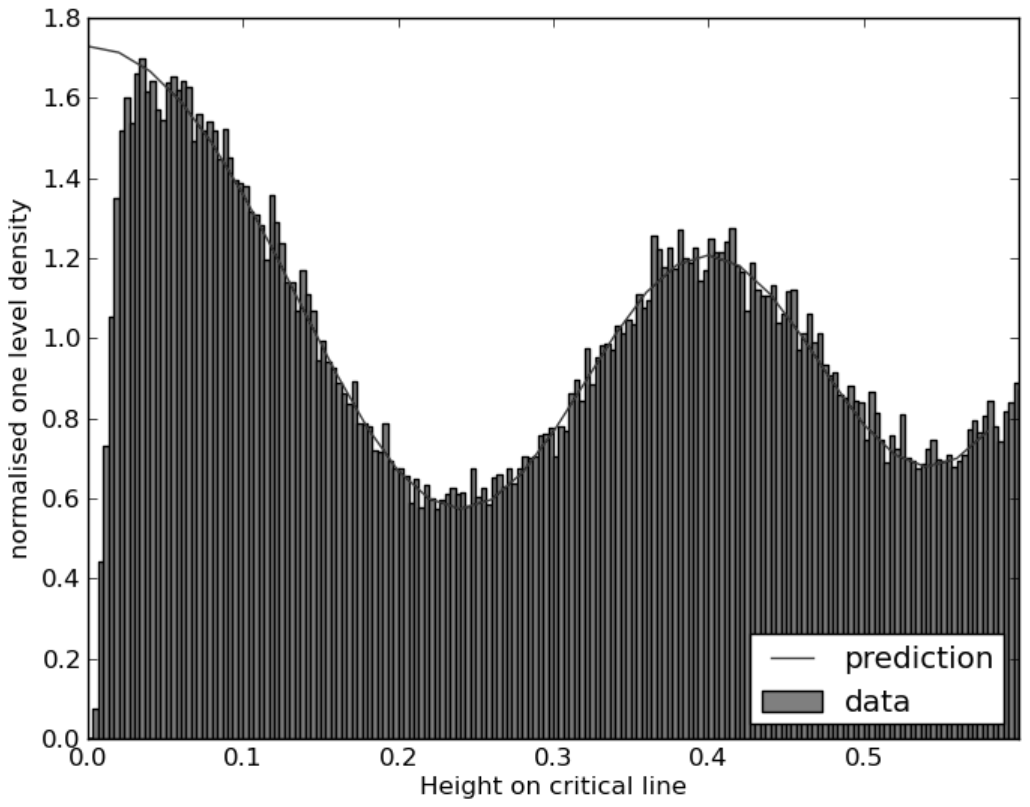}
\caption{One level density of zeros for even twists with $0<d\leq 400 \:000$ of the $L$-function associated with the elliptic curve $E11.a3$. We can see that the prediction due to the ratios conjecture as calculated in \cite{kn:hks} (the smooth curve, reproduced in (\ref{eq:riszero}) here) fails to capture the behaviour of the one level density close to the origin.}
\label{fig:E11_zero_repulsion}
\end{figure}

This corresponds to the one level density derived at Theorem 2.3 in \cite{kn:hks} (with slightly different notation for the arithmetic parts):
\begin{align}\label{eq:riszero}
\frac 1 {2 \pi} \int _{-\infty}^\infty f(\phi) \displaystyle \sum _{\substack{0 < d \leq X \\ \omega_E \chi_d(-M) = +1 }}  - \frac{X'}{X} (1/2 + \text i \phi, \chi_d)  + 2  \mathcal{U}_E(\mathcal{N}_d,0,\mathrm{i} \phi)   \mathrm{d} \phi   \nonumber \\
= \frac 1 {2 \pi} \int _{-\infty}^\infty f(\phi) \displaystyle \sum _{\substack{0 < d \leq X \\ \omega_E \chi_d(-M) = +1 }}    \left( 2 \log \left(\frac{ \sqrt{M} d}{2 \pi} \right) +  \Psi(1 + \mathrm{i} \phi )  +  \Psi(1 - \mathrm{i} \phi )  \right. \nonumber \\
+ 2 \left. \left[ - \tilde{A}_E(\mathrm{i} \phi,\mathrm{i} \phi)\frac{\zeta'}{\zeta} (1 + 2 \mathrm{i} \phi) + \tilde{A}^1_E(\mathrm{i} \phi) +\left(\frac{ \sqrt{M} d}{2 \pi}  \right) ^{- 2 \mathrm{i} \phi } \frac{\Gamma( 1 - \mathrm{i} \phi)}{\Gamma( 1 + \mathrm{i} \phi)} \zeta(1+2\text i \phi) \tilde{A}_E(-\mathrm{i} \phi,\mathrm{i} \phi)   \right]\right) \mathrm{d} \phi .
\end{align}
This is plotted in Figure \ref{fig:E11_zero_repulsion}.

%%%%%%%%%%%%%%%%%%%%%%%%%%%%%%%%%%%%%%%%%%
\subsection{The contribution from higher residues}
\begin{figure}[h!]
\centering
\includegraphics[width=10cm]{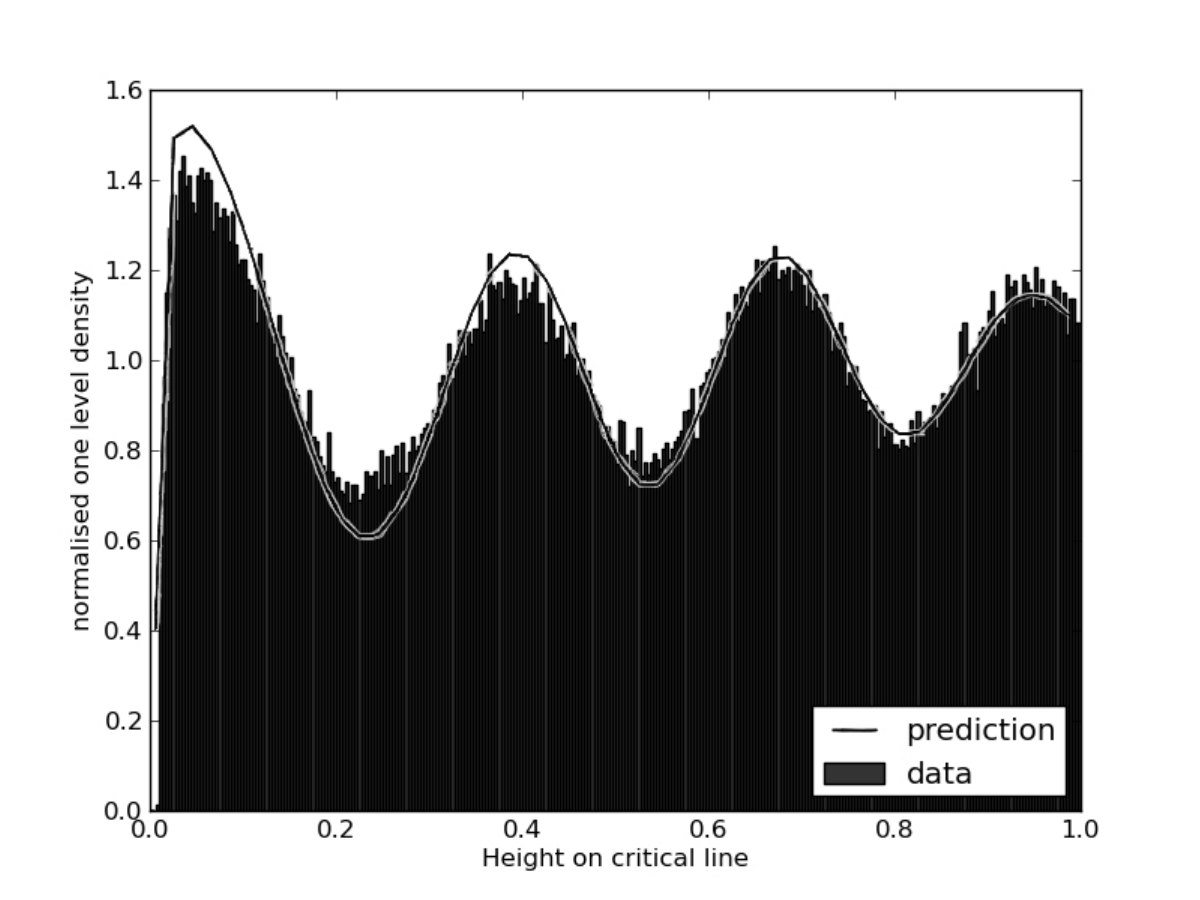}
\caption{One level density of zeros for even twists with $0 < d < 400000$ of the L-function associated with the elliptic curve $E11.a3$, with the first three residues from the prediction (\ref{eq:excised_density_prediction}).}
\label{fig:excised_density_prediction}
\end{figure}

The higher residues will not take such a neat form, but may be calculated using a computational algebra package, and the resulting one level density prediction using the first three residues is shown in Figure \ref{fig:excised_density_prediction}. Near the origin, this is a marked improvement on the prediction using only the first residue, i.e. the prediction from \cite{kn:hks} in Figure \ref{fig:E11_zero_repulsion}. Away from the origin, it is less good, but since we are only using the first three residues and we are only working to leading order term in $\mathcal{N}_X$, it is reasonable to suppose that if we calculated more residues and/or higher order terms in $\mathcal{N}_X$, then we would get even better agreement. This supports the hypothesis that information about the discretisation of the central values is lost in the error term of the ratios conjecture for finite conductor. 

\section*{Acknowledgements} The first author was supported by an EPSRC Doctoral Training Account. Thanks to Jon Keating for his help at the start of this project.

\section*{ Data availability Statement} The $L$-function data was obtained from Rubinstein's $L$-function calculator, which is included as standard in the open-source software Sage \begin{verbatim}https://www.sagemath.org\end{verbatim}.

%\bibliography{../../../bin/ref.bib}{}

\begin{thebibliography}{10}

\bibitem{kn:abprw14}
S.A. Altu{\v g}, S.~Bettin, I.~Petrow, Rishikesh, and I.~Whitehead.
\newblock A recursion formula for moments of derivatives of random matrix
  polynomials.
\newblock {\em Quart. J. Math.}, 65:1111--25, 2014.

\bibitem{kn:alvsna20}
E.~Alvarez and N.C. Snaith.
\newblock Moments of the logarithmic derivative of characteristic polynomials
  from {$SO(2N)$} and {$USp(2N)$}.
\newblock {\em Journal of Mathematical Physics}, 61(10), 2020.

\bibitem{kn:andbes}
J.C. Andrade and C.G. Best.
\newblock Joint moments of derivatives of characteristic polynomials of random
  symplectic and orthogonal matrices.
\newblock {\em J. Phys. A}, 57(20), 2024.
\newblock ar{X}iv:2312.04981.

\bibitem{kn:ashgross}
A.~Ash and R.~Gross.
\newblock {\em Curves, Counting and Number Theory}.
\newblock Princeton University Press, 2012.

\bibitem{kn:agkw}
T.~Assiotis, M.A. Gunes, J.P. Keating, and F.~Wei.
\newblock Exchangeable arrays and integrable systems for characteristic
  polynomials of random matrices.
\newblock ar{X}iv:2407.19233.

\bibitem{kn:ags22}
T.~Assiotis, M.A. Gunes, and A.~Soor.
\newblock Convergence and an explicit formula for the joint moments of the
  circular {J}acobi {$\beta$}-ensemble characteristic polynomial.
\newblock {\em Math. Phys. Anal. Geom.}, 25(15), 2022.

\bibitem{kn:akw22}
T.~Assiotis, J.P. Keating, and J.~Warren.
\newblock On the joint moments of the characteristic polynomials of random
  unitary matrices.
\newblock {\em IMRN}, 2022(18), 2022.

\bibitem{kn:bbbcprs}
E.C. Bailey, S.~Bettin, G.~Blower, J.B. Conrey, A.~Prokhorov, M.O. Rubinstein,
  and N.C. Snaith.
\newblock Mixed moments of characteristic polynomials of random unitary
  matrices.
\newblock {\em J. Math. Phys.}, 60(8), 2019.

\bibitem{kn:barhoumi}
Y.~Barhoumi-Andr{\'e}ani.
\newblock A new approach to the characteristic polynomial of a random unitary
  matrix.
\newblock ar{X}iv:2011.02465.

\bibitem{kn:barnes00}
E.W. Barnes.
\newblock The theory of the {$G$}-function.
\newblock {\em Q. J. Math.}, 31:264--314, 1900.

\bibitem{kn:basor_et_al18}
E.~Basor, P.~Bleher, R.~Buckingham, T.~Grava, A.~Its, E.~Its, and J.~Keating.
\newblock A representation of joint moments of {CUE} characteristic polynomials
  in terms of {P}ainlev{\'e} functions.
\newblock {\em Nonlinearity}, 32(10):4033--4078, 2019.
\newblock ar{X}iv:1811.00064.

\bibitem{kn:brezhik00}
E.~Br\'ezin and S.~Hikami.
\newblock Characteristic polynomials of random matrices.
\newblock {\em Comm. Math. Phys.}, 214:111--135, 2000.
\newblock ar{X}iv:math-ph/9910005.

\bibitem{kn:cfkrs}
J.B. Conrey, D.W. Farmer, J.P. Keating, M.O. Rubinstein, and N.C. Snaith.
\newblock Integral moments of ${L}$-functions.
\newblock {\em Proc. London Math. Soc.}, 91(1):33--104, 2005.
\newblock ar{X}iv:math.nt/0206018.

\bibitem{kn:cfz2}
J.B. Conrey, D.W. Farmer, and M.R. Zirnbauer.
\newblock Autocorrelation of ratios of {$L$}-functions.
\newblock {\em Comm. Number Theory and Physics}, 2(3):593--636, 2008.
\newblock ar{X}iv:0711.0718.

\bibitem{kn:ckrs05}
J.B. Conrey, J.P. Keating, M.O. Rubinstein, and N.C. Snaith.
\newblock Random matrix theory and the {F}ourier coefficients of half-integral
  weight forms.
\newblock {\em Experiment. Math.}, 15(1):67--82, 2006.
\newblock ar{X}iv:math.nt/0412083.

\bibitem{kn:crs06}
J.B. Conrey, M.O. Rubinstein, and N.C. Snaith.
\newblock Moments of the derivative of characteristic polynomials with an
  application to the {R}iemann zeta-function.
\newblock {\em Comm. Math. Phys.}, 267(3):611--629, 2006.
\newblock ar{X}iv:math.NT/0508378.

\bibitem{kn:consna06}
J.B. Conrey and N.C. Snaith.
\newblock Applications of the {$L$}-functions ratios conjectures.
\newblock {\em Proc. London Math. Soc.}, 94(3):594--646, 2007.
\newblock ar{X}iv:math.NT/0509480.

\bibitem{kn:consna08}
J.B. Conrey and N.C. Snaith.
\newblock Correlations of eigenvalues and {R}iemann zeros.
\newblock {\em Comm. Number Theory and Phyics}, 2(3):477--536, 2008.
\newblock ar{X}iv:0803.2795.

\bibitem{kn:consna07}
J.B. Conrey and N.C. Snaith.
\newblock Triple correlation of the {R}iemann zeros.
\newblock {\em Journal de Th{\'e}orie des Nombres de Bordeaux}, 20:61--106,
  2008.
\newblock ar{X}iv:math/0610495.

\bibitem{kn:coomorsna}
I.~A. Cooper, Patrick~W. Morris, and N.C. Snaith.
\newblock Beyond the excised ensemble: modelling elliptic curve {$L$}
  -functions with random matrices.
\newblock {\em Journal of Physics A: Mathematical and Theoretical}, 49(7),
  2016.

\bibitem{kn:dehaye10}
P.-O. Dehaye.
\newblock Joint moments of derivatives of characteristic polynomials.
\newblock {\em Alg. Number Theory}, 2(1):31--68, 2008.
\newblock ar{X}iv:math/0703440.

\bibitem{kn:dehaye10p}
P.-O. Dehaye.
\newblock A note on moments of derivatives of characteristic polynomials.
\newblock In {\em DMTCS Proceedings, 22nd International Conference on formal
  power series and algebraic combinatorics}, volume~AN. Discrete mathematics
  and Theoretical Computer Science, 2010.

\bibitem{kn:dhkms12}
E.~Due{\~n}ez, D.~K. Huynh, S.~J. Miller, J.~P. Keating, and N.~C. Snaith.
\newblock A random matrix model for elliptic curve {$L$}-functions of finite
  conductor.
\newblock {\em J. Phys. A}, 45(11), 2012.
\newblock arXiv:1107.4426.

\bibitem{kn:fazzari}
A.~Fazzari.
\newblock On the joint second moment of zeta and its logarithmic derivative.
\newblock arXiv:2310.15918.

\bibitem{kn:forrester22}
P.J. Forrester.
\newblock Joint moments of a characteristic polynomial and its derivative for
  the circular {$\beta$}-ensemble.
\newblock {\em Probability and mathematical physics}, 3(1), 2022.

\bibitem{kn:heineman29}
E.R. Heineman.
\newblock Generalized {V}andermonde determinants.
\newblock {\em Trans. AMS}, 31(3):464--476, 1929.

\bibitem{kn:hug01}
C.P. Hughes.
\newblock {\em On the characteristic polynomial of a random unitary matrix and
  the {R}iemann zeta function}.
\newblock PhD thesis, University of Bristol, 2001.

\bibitem{kn:hks}
D.K. Huynh, J.P. Keating, and N.C. Snaith.
\newblock Lower order terms for the one-level density of elliptic curve
  {$L$}-functions.
\newblock {\em J. Number Theory}, 129:2883--2902, 2009.
\newblock ar{X}iv:0811.2304.

\bibitem{kn:katzsarnak99a}
N.M. Katz and P.~Sarnak.
\newblock {\em Random Matrices, Frobenius Eigenvalues and Monodromy}.
\newblock American Mathematical Society Colloquium Publications, 45. American
  Mathematical Society, Providence, Rhode Island, 1999.

\bibitem{kn:katzsarnak99b}
N.M. Katz and P.~Sarnak.
\newblock Zeros of zeta functions and symmetry.
\newblock {\em Bull. Amer. Math. Soc.}, 36:1--26, 1999.

\bibitem{kn:keasna00b}
J.P. Keating and N.C. Snaith.
\newblock Random matrix theory and ${L}$-functions at $s=1/2$.
\newblock {\em Comm. Math. Phys}, 214:91--110, 2000.

\bibitem{kn:keawei1}
J.P. Keating and F.~Wei.
\newblock Joint moments of higher order derivatives of {CUE} characteristic
  polynomials {I}: asymptotic formulae.
\newblock {\em IMRN}, 12, 2024.
\newblock ar{X}iv:2307.01625.

\bibitem{kn:keawei2}
J.P. Keating and F.~Wei.
\newblock Joint moments of higher order derivatives of {CUE} characteristic
  polynomials {II}: structures, recursive relations and applications.
\newblock {\em Nonlinearity}, 37(8), 2024.
\newblock ar{X}iv:2307.02831.

\bibitem{kn:kohzag81}
W.~Kohnen and D.~Zagier.
\newblock Values of ${L}$-series of modular forms at the center of the critical
  strip.
\newblock {\em Invent. Math.}, 64:175--198, 1981.

\bibitem{kn:massna16}
A.M. Mason and N.C.Snaith.
\newblock Orthogonal and symplectic {$n$}-level densities.
\newblock {\em Memoirs of the AMS}, 251(1194), 2018.
\newblock ar{X}iv:1509.05250.

\bibitem{kn:baruch_mao}
E.~{Moshe Baruch} and Z.~{Mao}.
\newblock {Central values of automorphic $L$-functions}.
\newblock {\em Geom. Func. Anal.}, 17:333--384, 2007.

\bibitem{kn:rubsil02}
K.~Rubin and A.~Silverberg.
\newblock Ranks of elliptic curves.
\newblock {\em Bull. Amer. Math. Soc.}, 39(4):455--74, 2002.

\bibitem{kn:simmwei}
N.~Simm and F.~Wei.
\newblock Private communication.

\bibitem{kn:waldspurger81}
J.-L. Waldspurger.
\newblock Sur les coefficients de {F}ourier des formes modulaires de poids
  demi-entier.
\newblock {\em J. Math. Pures Appl.}, 60(9):375--484, 1981.

\bibitem{kn:winn12}
B.~Winn.
\newblock Derivative moments for characteristic polynomials from the {CUE}.
\newblock {\em Commun. Math. Phys.}, 315:531--562, 2012.

\end{thebibliography}
%\bibliographystyle{plain}

\end{document}